\documentclass[12pt,a4paper]{article}

\usepackage{epsf}
\usepackage{amsmath}
\usepackage{bbm}
\usepackage{rotate}
\usepackage[hypertex]{hyperref}
\renewcommand{\baselinestretch}{1.5}
\makeatletter \@addtoreset{equation}{section} \makeatother

\newcommand{\be}{\begin{equation}}
\newcommand{\ee}{\end{equation}}
\newcommand{\bea}{\begin{eqnarray}}
\newcommand{\eea}{\end{eqnarray}}
\newcommand{\cN}{\mathcal{N}}
\newcommand{\cL}{\mathcal{L}}

\newcommand{\cV}{\mathcal{V}}
\newcommand{\pV}{\mathbbm{V}}
\newcommand{\cD}{\mathcal{D}}

\newcommand{\cW}{\mathcal{W}}
\newcommand{\cM}{\mathcal{M}}

\newcommand{\cQ}{{\mathcal{Q}}}
\newcommand{\C}{\mathbbm{C}}

\newcommand{\Rom}{\mathbbm{R}}

\newcommand{\unit}{{\mathbbm{1}}}
\newcommand{\imag}{i}
\newcommand{\rmd}{\mbox{\rm{d}}}

\newcommand{\ub}{\bar{u}}
\newcommand{\vb}{\bar{v}}
\newcommand{\jb}{\bar{\jmath} }
\newcommand{\ib}{\bar{\imath} }
\newcommand{\kb}{\bar{k}}
\newcommand{\lb}{\bar{l}}

\newcommand{\kh}{\hat{k}}

\newcommand{\mscr}[1]{\mbox{\scriptsize #1}}
\newcommand{\ft}[2]{{\textstyle\frac{#1}{#2}}}
\begin{document}
\begin{titlepage}

\begin{center}
\renewcommand{\baselinestretch}{1} \large\normalsize
\hfill hep-th/0410272\\
\hfill FSU-TPI-07/04 \\
\hfill ITP-UU-04/36 \\
\hfill SPIN-04/19 \\
\renewcommand{\baselinestretch}{1.5} \large\normalsize
\renewcommand{\thefootnote}{\fnsymbol{footnote}}
\vskip 1cm {\large \bf  Effective Supergravity Actions \\ for Conifold
  Transitions}\footnote{Work supported by the 'Schwerpunktprogramm Stringtheorie' of the DFG.}
\renewcommand{\thefootnote}{\arabic{footnote}}

\vskip 0.7cm

{\bf Thomas Mohaupt$^a$ and Frank Saueressig$^b$}  \\[3ex]
\renewcommand{\baselinestretch}{1} \large\normalsize
$^a${\em Institute of Theoretical Physics,
Friedrich-Schiller-University 
Jena, \\ 
 Max-Wien-Platz 1, D-07743 Jena, Germany} \\
{\tt Thomas.Mohaupt@uni-jena.de}\\[2.5ex]
$^b${\em Institute of Theoretical Physics} and {\em Spinoza Institute, \\
Utrecht University, 3508 TD Utrecht, The Netherlands} \\
{\tt F.S.Saueressig@phys.uu.nl} \\ 
\renewcommand{\baselinestretch}{1.5} \large\normalsize
\end{center}
\vskip 0.7cm

\begin{center} {\bf ABSTRACT } \end{center}
\noindent
We construct gauged supergravity actions which 
describe the dynamics of M-theory on a Calabi-Yau threefold
in the vicinity of a conifold transition. The actions explicitly
include $N$ charged hypermultiplets descending from wrapped
M2-branes
which become massless
at the conifold point. While the vector multiplet sector
can be treated exactly, we approximate the 
hypermultiplet sector by the non-compact Wolf spaces $X(1+N)$. 
The effective action is then uniquely determined by the charges
of the wrapped M2-branes.
%
\noindent

\end{titlepage}

\tableofcontents

\setcounter{footnote}{0}

\begin{section}{Introduction}
\end{section}

Conifold transitions are the most studied example of topology 
change in string theory. But when they first came into the focus
of physicists working on Calabi-Yau compactifications \cite{connected2,
connected,Candelas2,Candelas1}
they posed a considerable problem. The mathematics 
was well understood by that time. As explained in detail
in \cite{Huebsch}, conifold transitions occur when 
a singular Calabi-Yau space (in the following: CY space) admits 
two different types of 
smoothings of its singular points: by either replacing each
singularity by a two-sphere (the small resolution) or by replacing
it by  a three-sphere (the deformation).\footnote{A more detailed
account is given in Section 2.} To make a conifold transition 
one tunes $N-r$ K\"ahler moduli of a smooth 
CY space $X$ with Hodge numbers $(h^{1,1}, h^{1,2})$ 
such that $N$ two-spheres, $N-r$ of which are independent
in homology, 
shrink to zero volume. 
The resulting singular CY space $\hat{X}$ is then deformed
into a new smooth CY space $\tilde{X}$ by turning on $r$ new complex
structure moduli, which control the sizes of the $N$ three-spheres,
$r$ of which are homologically
independent. The space $\tilde{X}$ has Hodge
numbers $\tilde{h}^{1,1} = h^{1,1} - (N-r)$ and $\tilde{h}^{1,2} = 
h^{1,2} + r$. Many CY spaces are related this way, and a mathematical
conjecture, known as Reid's conjecture (or Reid's fantasy) even claims
that all CY manifolds form a single connected web \cite{Reid}. This conjecture
has gained much support from string theory, starting from \cite{connected2,
connected,Candelas2,Candelas1}.

Despite their geometrical elegance, conifold transitions originally provided
a problem for string theory, because the metric of the moduli space
becomes singular at conifold points,\footnote{In this paper,
`conifold point' always refers to the point in moduli space where
a CY space becomes singular. The singular points of the CY space
are referred to as `nodes' or `singular points.'} 
and therefore there is no
well defined  world-sheet CFT corresponding to the singular CY space 
$\hat{X}$. Likewise, the low energy effective actions (LEEA) $S$
and $\tilde{S}$, which contain the massless modes of string compactifications
on the spaces $X$ and $\tilde{X}$, respectively, 
become singular. Since the conifold points 
are at finite distance with respect to the metric on moduli space,
it seemed that well defined string backgrounds could develop 
into singular ones within finite time. This changed when the 
role of branes in string theory was appreciated. As pointed out
in \cite{Str1}, the low energy effective action (LEEA)
$\tilde{S}$ corresponding to the compactification 
of the type IIB string on $\tilde{X}$ develops a singularity
at the conifold point, which is precisely of the form one
gets when integrating out $N$ charged hypermultiplets.\footnote{We
always denote a CY space which becomes 
singular through tuning complex structure moduli by $\tilde{X}$,
while we denote a space which becomes singular through
tuning K\"ahler moduli by $X$. The singular space itself
is denoted $\hat{X}$. The corresponding LEEA are
$\tilde{S}$, $S$, and $\hat{S}$, respectively.}
Moreover, type IIB
string theory indeed contains $N$ charged hypermultiplets which become
massless at the conifold point, namely the winding states of D3-branes
around the $N$ three-spheres which are contracted. The conifold 
transition from $\tilde{X}$ to $X$ can then be interpreted as the
transition from the Coulomb branch to the Higgs branch of an
Abelian gauge theory \cite{Str2}. Conifold transitions work
analogously in type IIA string theory and M-theory. Here, the
extra hypermultiplets on the Coulomb branch come from D2-branes
and M2-branes, respectively, which wrap two-spheres inside 
$X$. Thus there is at least a   
consistent `macroscopic' description of conifold transitions
in terms of four-dimensional,
or, in the M-theory case, five-dimensional supergravity. 
Moreover it is clear that there must exist an extended
LEEA $\hat{S}$ which is regular at the conifold point and
includes 
the `transition states'\footnote{The term `transition states'
was coined in \cite{flop}. In this paper we will refer
to LEEA which explicitly contain transition states as 
`extended LEEA.' In \cite{FS3,FS4} this is called the `in-picture
LEEA.'} 
as dynamical degrees of freedom. 
The actions $S$ and $\tilde{S}$  corresponding to
compactifications on the smooth manifolds $X$ and $\tilde{X}$,
respectively, can be interpreted as arising from integrating
out the transition states. While the properties of some terms
in $\hat{S}$ were discussed in \cite{Str1,Str2}, the full 
action $\hat{S}$ has not been worked out so far. But since
all the light modes of a system enter into its low energy dynamics,
it is desirable to know $\hat{S}$ explicitly, 
in order to study the behavior of string compactifications close
to a conifold transition. Similar remarks
apply to other situations in string compactifications
where additional light modes occur at special points in the 
moduli space, in particular to flop transitions in CY compactifications
and to compactifications on self-dual circles or tori. 

In this paper we continue the work of \cite{MohZag,flop,LouMohZag,FS3,FS4} 
on such extended LEEA
and construct $\hat{S}$ for  conifold transitions in M-theory on CY 
threefolds. While the
vector multiplet sector of $\hat{S}$ can be found exactly for
arbitrary conifold transitions, it is currently out of reach
to derive the metric on the hypermultiplet manifold from string or M-theory.\footnote{
We refer to \cite{Asp} for a review of hypermultiplet manifolds
in string compactifications.} Therefore
we take the simplest family of hypermultiplet
manifolds, the non-compact Wolf spaces $X(1+N)$, to model the 
universal hypermultiplet together with the $N$ hypermultiplets which
become massless at the conifold point. M-theory specifies
the charges carried by these transition states, and for 
transitions with $r=1$  and arbitrary $N$ we show 
explicitly that  this determines a unique action $\hat{S}$. We further
show that this action has all the properties 
needed to describe a conifold transition. In particular 
the model has a non-vanishing scalar potential whose flat directions
correspond to a Coulomb and a Higgs branch.  Therefore we 
interprete our model as the leading-order supergravity 
approximation of the full M-theory LEEA. The physical properties
of these actions are investigated in a companion paper \cite{CC}. 
There, we present numerical solutions which undergo dynamical
conifold transitions, and we
show that the inclusion of the transition states has important
consequences for string cosmology and moduli stabilization.

This paper is organized as follows: in Section 2 
we collect the necessary background material on 
five-dimensional supergravity, compactifications on 
CY spaces, and conifold transitions. In Section 3 we 
construct a model with the minimal field content needed
to describe a conifold transition. This prepares the stage
for a more general model which corresponds to conifold 
transitions with $r=1$ and arbitrary $N$ in Section 4. 
We present our conclusions in Section 5. Some technical
details concerning the vector multiplet manifolds 
are relegated to the Appendices A and B. In Appendix C
we present a particular conifold transition to give 
the interested reader a more detailed account of the
geometrical aspects.

\begin{section}{Background material}
\begin{subsection}{Five-dimensional supergravity}
The LEEA of eleven-dimensional supergravity compactified on a smooth 
CY threefold $X$ with Hodge numbers $h^{p,q}$ is given by five-dimensional supergravity coupled to 
$n_V = h^{1,1}-1$ Abelian vector and $n_N = h^{2,1}+1 $ neutral 
hypermultiplets. 
When explicitly including the 
transition states arising in a conifold transition we obtain 
$\delta n_H$ additional charged hypermultiplets. 

The LEEA which includes these states is given by the general $\cN = 2, D = 5$
gauged supergravity action \cite{5dSuGra}, specialized to the case of
$n_V$ vector and $n_H = n_N + \delta n_H$ 
hypermultiplets. 
Anticipating the results of the subsequent sections, we limit ourselves 
to the case of Abelian gaugings.
The bosonic matter content of this theory consists of the graviton $e_\mu^{~a}$, \mbox{$n_V+1$} vector fields $A^I_\mu$ with field 
strength $F_{\mu \nu}^I = \partial_\mu A^I_\nu - \partial_\nu A^I_\mu$, $n_V$ real vector multiplet scalars $\phi^x$, and $4 
n_H$ real hypermultiplet scalars $q^X$. The bosonic part of the Lagrangian reads: 
\bea\label{2.1}
\nonumber \sqrt{-g}^{-1} \cL^{\cN = 2}_{\rm bosonic} & = &  - \frac{1}{2} R - \frac{1}{4} a_{IJ} F^I_{\mu\nu} F^{J~\mu\nu} 
\\ 
&& - \frac{1}{2} g_{XY} \cD_{\mu} q^X \cD^{\mu} q^Y - \frac{1}{2} g_{xy} \cD_{\mu} \phi^x \cD^{\mu} \phi^y \\ 
\nonumber && + \frac{1}{6 \sqrt{6}} C_{IJK} \sqrt{-g}^{-1} \epsilon^{\mu\nu\rho\sigma\tau} F^I_{\mu \nu} F^J_{\rho \sigma} 
A^K_\tau - {\rm g}^2 \pV(\phi, q) \, .  \eea
The scalars $\phi^x$ and $q^X$ parameterize a 
very special real manifold $\cM_{\mscr{VM}}$ \cite{VSG} and 
a quaternion-K\"ahler
%
%
manifold $\cM_{\mscr{HM}}$ \cite{BW}. The quaternion-K\"ahler condition 
implies that $\cM_{\mscr{HM}}$ is an Einstein manifold.
A ghost-free action requires in addition that 
the Ricci scalar satisfies ${\cal R}= -4 n_H (n_H +2)$, as 
is readily seen using the superconformal calculus \cite{CHM}.

The vector multiplet sector is determined by the completely symmetric tensor $C_{IJK}$, appearing 
in the Chern-Simons term. This tensor is used to define a real homogeneous
cubic polynomial
\be\label{2.3}
{\cal V}(h) = C_{IJK} \, h^I \, h^J \, h^K \, 
\ee
in $n_V + 1$ real variables $h^I$. The $n_V$-dimensional 
manifold ${\cal M}_{\mscr{VM}}$ 
is obtained by restricting this polynomial to the hypersurface 
\be\label{2.4}
{\cal V}(\phi) = C_{IJK} \, h^I(\phi) \, h^J(\phi) \, h^K(\phi) = 1 \, .
\ee
The coefficients $a_{IJ}$ appearing in the kinetic term of the vector field strength are given by
\be\label{2.5}
\begin{split}
a_{IJ}(h) & := - \, \left. \frac{1}{3} \, \frac{\partial}{\partial h^I} \, \frac{\partial}{\partial h^J} \, \ln {\cal V}(h) 
\right|_{ {\cal V} = 1}  \\
& = -2 \, C_{IJK} \, h^K + 3 \, C_{IKL} \, C_{JMN} \, h^K h^L h^M h^N \, .
\end{split}
\ee
Defining
\be\label{2.6}
h^I_x := - \, \sqrt{\frac{3}{2}} \, \frac{\partial}{\partial \phi^x} \, h^I(\phi) \, , \quad h_I := C_{IJK} h^J h^K \, ,
\ee
the metric on $\cM_{\mscr{VM}}$ is proportional to the pullback\footnote{The 
matrix $a_{IJ}$
can be interpreted as a metric on an $(n_V+1)$-dimensional 
space, into which ${\cal M}_{\mscr{VM}}$
is immersed by (\ref{2.4}).}
of $a_{IJ}$,
\be\label{2.7}
g_{xy}(\phi) := h^I_x \, h^J_y \, a_{IJ} \,.
\ee

The hypermultiplet scalars $q^X$ parameterize a quaternion-K\"ahler manifold of dimension $\dim_{\Rom}(\cM_{\mscr{HM}}) = 4 n_H$. 
For $n_H > 1$ such manifolds are characterized by their holonomy group,
\be\label{2.7a}
Hol(\cM_{\mscr{HM}}) = SU(2) \cdot USp(2 n_H) \, ,
\ee
 while in the case $n_H = 1$ they are defined as Einstein spaces
with self-dual Weyl curvature. In general quaternion-K{\"a}hler manifolds 
need not be K\"ahler manifolds or complex manifolds. However, they
are quaternionic manifolds, i.e., locally there exists 
a triplet $J^r$ of almost complex structures, which satisfy 
the quaternionic algebra
\be\label{2.7b}
J^r \, J^s = - \delta^{rs} + \epsilon^{rst} \, J^t \, , 
\quad r,s,t = 1,2,3 \, .
\ee
%

%
In our construction of extended LEEA for conifold transitions we will
use the non-compact Wolf spaces
\be\label{1.1}
X(1+N) = \frac{U(1+N,2)}{U(1+N) \times U(2)} \, ,
\ee
as hypermultiplet manifolds.
For $N = 0$ this series contains the universal hypermultiplet $\frac{U(1,2)}{U(1) \times U(2)}$. To utilize these spaces in the construction of gauged supergravity actions, we need to know the metric, Killing vectors and the associated $SU(2)$-triplets of moment maps of $X(1+N)$. 
Using the superconformal quotient construction \cite{SCQ,SCgauging,SCQ2}, 
these geometrical data have been obtained in \cite{FS3}. As it will turn out the resulting parameterization is very useful for our construction and we now briefly summarize the relevant results. 

We first introduce complex coordinates $v^i, u_i$, $i = 1, \ldots, N+1$ with
respect to the complex structure $J^3$.\footnote{While for a general
quaternion-K\"ahler manifold $J^3$ does not need to be integrable,
it happens to be integrable for $X(1+N)$. In the construction 
used in \cite{SCQ,SCgauging,SCQ2,FS3}
this is manifest, because $X(1+N)$ inherits a $J^3$-complex 
coordinate system from the associated hyper-K\"ahler cone. 
Note, however, that the associated fundamental form
is not closed, and, therefore, the quaternion-K\"ahler 
metric is not K\"ahler with 
respect to this complex structure \cite{SCQ}.}
The hypermultiplets of $X(N+1)$ are described by the coordinates $v^i,u_i$ with fixed index $i$. Using these coordinates we see that the metric on $X(N+1)$ 
is hermitian with respect to $J^3$. Its components read\footnote{Note that
  there is a relative minus sign between the metric obtained from the superconformal quotient construction and the conventions of the supergravity Lagrangian (\ref{2.1}). The results given here are adapted to the supergravity Lagrangian.}
\bea\label{3.31}
\nonumber g_{u_i \ub_{ \jb }}  & = & - \, \frac{1}{2 \phi_-}  \left( \eta^{i \jb} + \vb^{\jb} \, v^i \right) 
+ \frac{1}{2 \phi_-^2}
\left( \eta^{\jb l} u_l + \vb^{\jb} \left( v^l u_l \right) \right)  
\big( \eta^{i \lb} \ub_{\lb} + v^i \big( \vb^{\lb} \ub_{\lb} \big) \big) \, , \\
g_{\vb^{\ib} u_{j}} & = & - \, \frac{1}{2 \phi_-^2} \, \left( \ub_{\ib} v^j \, \left( 1 + \eta^{k \lb} u_k \ub_{\lb} \right) - 
\ub_{\ib} \eta^{j \lb} \ub_{\lb} \left( v^l u_l \right) \right)  \, , \\
\nonumber g_{v^i \vb^{\jb}} & = & 
- \, \frac{1}{2 \phi_+} \, \eta_{i \jb} 
+ \frac{1}{2 \phi_+^2} \, \big( \eta_{i \lb} \, \vb^{\lb} \big) \, \big( \eta_{\jb l} \, v^{l} \big) \, + \, \frac{1}{\phi_+ 
\phi_-} \, u_i \ub_{\jb} \\
\nonumber && - \, \frac{1}{2 \phi_-} \, u_i \, \ub_{\jb} \, + \, \frac{1}{2 \phi_-^2} \, u_{i} \, \ub_{\jb} \, \big( v^l \, 
u_l \big) \big( \vb^{\lb} \ub_{\lb} \big) \, .
\eea
Here $\eta_{i \bar{\jmath}} := {\rm diag} \left[ -1, \ldots -1 \right]$ is minus the $(n-1) \times (n-1)$-dimensional unit matrix, $\eta^{i \bar{\jmath}}$ denotes its inverse, and $\phi_+$ and $\phi_-$ are given by
\be\label{3.25}
 \phi_+ := 1 + \eta_{i \bar{\jmath}} \, v^i \bar{v}^{\bar{\jmath}}  \, , \quad 
\phi_- :=  1 + \eta^{i \bar{\jmath}} \, u_i \bar{u}_{\bar{\jmath}} + \left( v^i u_i \right) \left( \bar{v}^{\bar{\imath}} \, 
\bar{u}_{\bar{\imath}} \right)  \, .
\ee

The triholomorphic Killing vectors of the metric (\ref{3.31}) are completely determined by the generators $\left[ \, t \,  \right]^{J}_{~I}$, $I  = i, n , n+1$ of the Lie algebra $su(N+1,2)$ of the isometry group of (\ref{1.1}). Explicitly they read
\bea\label{3.42}
\nonumber \kh^{v^i} & = & \imag \, t^{i}_{~j} \, v^j + \imag \, t^{i}_{~n+1} - \imag \, v^{i} \, t^{n+1}_{~~~j} \, v^{j} - 
\imag \, v^{i} \, t^{n+1}_{~~~n+1} - \, \frac{ k^{\zeta} }{2} \left( \eta^{i \jb} \ub_{\jb} + v^{i} \, \vb^{\kb} \ub_{\kb} 
\right) \, ,\\
\kh^{u_i} & = & \imag \, u_i \, \left( t^{n+1}_{~~~j} v^j + t^{n+1}_{~~~n+1} \right) - \imag \, t^{j}_{~i} \, u_{j} - \imag 
\, t^{n}_{~i} + \imag \, t^{n+1}_{~~~i} \, \left( v^j \, u_j \right) \, \\
\nonumber && \qquad \qquad  - 2 u_i \, k_z + \frac{\phi_-}{2 \phi_+} \, k^{\zeta} \, \eta_{i \jb} \, \vb^{\jb} \, ,
\eea
with $k^{\zeta}$ and $k_z$ given by
\be\label{3.35}
\begin{split}
k^{\zeta} & = 2 \, \imag \, \left( t^n_{~i} \, v^i + t^n_{~n+1} \right) \, \quad \mbox{and} \\
k_z & = \frac{\imag}{2} \left( t^{n+1}_{~~~i} v^i + t^{n+1}_{~~~n+1} - t^{i}_{~n} \, u_i - t^n_{~n} + t^{n+1}_{~~~n} \, v^i 
\, u_i \right) \, .
\end{split}
\ee
The components of the Killing vectors with respect to $\partial_{\vb^{\ib}}$ and $\partial_{\ub_{\ib}}$ are obtained from eq. (\ref{3.42}) by complex conjugation.

The $SU(2)$-triplet of moment maps associated with these Killing vectors is
\bea\label{3.45}
\hat{\mu}^3 & = & - \, \frac{1}{2 \phi_+} \left\{ \vb^{\ib} \, \eta_{\ib j} \, t^j_{~k} \, v^k + t^{n+1}_{~~~i} \, v^i + 
\vb^{\jb} \, \eta_{\jb i} \, t^i_{~n+1} + t^{n+1}_{~~~n+1} \right\} \\ \nonumber
&&  + \frac{1}{2 \phi_-} 
\big\{  
u_i \, t^i_{~j} \, \eta^{j \kb} \, \ub_{\kb} + u_i \, t^{i}_{~n} + t^n_{~i} \, \eta^{i \jb} \, \ub_{\jb} + t^n_{~n} 
- \left( u_i \, t^i_{~n+1} + t^n_{~n+1}\right) \left( \ub_{\ib} \vb^{\ib} \right) 
\\ \nonumber
&& \qquad \qquad 
- \left( t^{n+1}_{~~~i} \, \eta^{i \jb} \, \ub_{\jb} + t^{n+1}_{~~~n} \right)
 \, \left( u_k v^k \right) 
+ t^{n+1}_{~~~n+1} \, \left( u_i \, v^i \right) 
\, \left( \ub_{\jb} \vb^{\jb} \right)
\big\}  \, , 
\\ \nonumber
\hat{\mu}^+ & = & - \, \frac{\imag}{2 \phi_+^{1/2} \phi_-^{1/2}} \, \big\{ u_i \, t^i_{~j} \, v^j + t^n_{~j} \, v^j - 
t^{n+1}_{~~~i} \, v^i \left( u_j \, v^j \right) + u_i \, t^i_{~n+1} \\ \nonumber
&& \qquad \qquad \qquad + t^n_{~n+1} - t^{n+1}_{~~~n+1} \left(u_i \, v^i \right) \big\} \, .
\eea
These are related to the quantities $P^r_I$ appearing in the scalar potential $\pV$ of the Lagrangian (\ref{2.1}) by
\be\label{3.51}
P^1_I = \frac{1}{2} \left(\hat{\mu}^+ + \hat{\mu}^-  \right) \, , \quad P^2_I = - \, \frac{\imag}{2} \left( \hat{\mu}^+ - \hat{\mu}^- \right) \, , \quad P^3_I = \frac{1}{2} \hat{\mu}^3 \, , 
\ee
where $\hat{\mu}^-$ is the complex conjugate of $\hat{\mu}^+$.

Since a conifold transition involves charged hypermultiplets,
we must specify an appropriate gauging of the Lagrangian (\ref{2.1}).
The scalars take values in a Riemannian manifold, and the gauge
group must operate on them as a subgroup of the isometry group
in order to have an action which is both gauge invariant and
invariant with respect to reparameterizations of the scalar manifold. 
As part of the gauging procedure we need to   
covariantize the derivatives of the scalar fields with 
respect to isometries 
of the vector or hypermultiplet target manifolds,
\be\label{2.2}
\cD_{\mu} q^X := \partial_{\mu} q^X + {\rm g} A^I_\mu K^X_{I}(q) \, , \quad 
\cD_{\mu} \phi^x := \partial_{\mu} \phi^x + {\rm g} A^I_\mu K^x_{I}(\phi) \, .
\ee
Here the  $ K^X_{I}(q)$ and $ K^x_{I}(\phi)$ are the Killing vectors of the gauged isometries in the hypermultiplet and 
vector multiplet scalar manifold, respectively. As an important consequence of the gauging we now have a non-trivial scalar potential $\pV(\phi, q)$. Since 
we have both vector and hypermultiplets but no tensor multiplets, 
and since the vector multiplets remain uncharged,
the scalar  potential is determined by the  gauging of the hypermultiplet 
isometries. In order to 
write down $\pV$ 
explicitly, we define:
\be\label{2.19}
P^r(\phi,q) := h^I(\phi) P^r_I(q) \, , \quad P^r_x(\phi,q) := h^I_x(\phi) P^r_I(q) \, , \quad K^X(\phi,q) := h^I(\phi) K^X_I(q) \, .
\ee
Here $h^I(\phi)$ are the scalars (\ref{2.3}), which are associated to the 
gauge fields $A^I_\mu$, $K_I^X(q)$ denotes the Killing 
vector of the hypermultiplet isometry for which $A^I_\mu$ serves as a gauge connection, and $P^r_I$ is its associated $SU(2)$ triplet of moment maps. 
%
%
%
The scalar potential takes the form
\be\label{2.20}
\pV(\phi, q) = -4 P^r P^r + 2 g^{xy} P^r_x P^r_y  + \frac{3}{4} g_{XY} 
K^X K^Y \;.
\ee

To discuss the vacuum structure of this potential, it is further useful to introduce the real `superpotential' \cite{UHM}
\be\label{2.21}
\cW := \sqrt{\frac{2}{3} \, P^r \, P^r } \, ,
\ee
which can be read off from the supersymmetry variations of the gravitino. Under the condition that the phase $Q^r$, which is  defined by
\be\label{2.22}
P^r = \sqrt{\frac{3}{2}} \, \cW \, Q^r \, , \quad Q^r \, Q^r = 1 \, ,
\ee
is independent of the vector multiplet scalars, $\partial_x Q^r = 0$, the scalar potential (\ref{2.20}) can be rewritten in terms of $\cW$:
\be\label{2.23}
\pV(\phi, q) = -6 \cW^2 + \frac{9}{2} g^{\Lambda \Sigma} \partial_{\Lambda} \cW \partial_{\Sigma} \cW \, .
\ee
Here $\phi^\Lambda$, $\Lambda, \Sigma = 1, \ldots , n_V + 4 n_H$ denotes the 
combined set of vector and hypermultiplet 
scalar fields and $g^{\Lambda \Sigma}$ is the direct sum of the  vector and hypermultiplet inverse metrics,
\be\label{4.14}
g^{\Lambda \Sigma}(\phi, q) := g^{XY}(q) \oplus g^{xy}(\phi) \, .
\ee
The `stability form' (\ref{2.23}) is useful, because
it is sufficient to guarantee the gravitational stability of the
theory \cite{stabil}. 

It was further shown in \cite{UHM,Vicente} that the minima of $\cW$ are given by the solution of the algebraic equations\footnote{For analogous relations in 
$\cN = 2, D = 4$ super-Yang-Mills theory see \cite{AP}.}
\be\label{2.24}
h_I\left(\phi_{c} \right) \, P^r \left( \phi_{c}, q_{c}\right) =  P^r_I \left( q_{c} \right) \quad \mbox{and} \quad 
K^X(\phi_{c}, q_{c}) = 0 \, .
\ee 
Here $\phi_c$ and $q_c$ denote the restrictions of the scalar fields
$\phi$ and $q$ to the critical
submanifolds of the vector and hypermultiplet scalar manifolds where the 
conditions (\ref{2.24}) are satisfied. These 
submanifolds correspond to the supersymmetric vacua of the theory. Indeed, it is straightforward to check that the solutions of the equations (\ref{2.24}) are 
critical points of the potential (\ref{2.20}) with ${\mathbbm V}|_{\phi_c,q_c} \le 0$, implying that the resulting vacua are Minkowski or anti-de Sitter. 
Further, $\partial_x Q^r = 0$ is always satisfied on the critical 
submanifolds, so that the restriction of ${\mathbbm V}$ to these
subspaces can always be
rewritten in the stability form (\ref{2.23}). 

In the following discussion, two types of solutions to the equations (\ref{2.24}) will play a role. First, the conditions (\ref{2.24}) may fix (some of) the hypermultiplet scalars, while the vector multiplet scalars are unconstrained. 
This is the Coulomb branch. Second there are solutions where (some of) the vector
multiplet scalars are fixed while the hypermultiplet scalars are free to take
values in a submanifold of $\cM_{\rm HM}$. This corresponds to the 
Higgs branch.
In between these extreme cases there can also be mixed branches.

%
%
%
%
%
%
%
\end{subsection}

\begin{subsection}{Calabi-Yau compactifications}
When compactifying eleven-dimensional supergravity on a smooth CY threefold $X$ \cite{CY_1}, one obtains a five-dimensional ungauged supergravity action, i.e, 
all 
fields are neutral under the gauge group $U(1)^{n_V + 1}$ and
there is no scalar potential.
In this case the objects introduced above acquire the following 
geometrical interpretation: the vector multiplet scalars encode the deformations of the 
K\"ahler class of $X$ at fixed total volume, while the hypermultiplet
scalars parameterize the volume of $X$, deformations of its 
complex structure, and deformations of the three-form gauge field
of eleven-dimensional supergravity.
The hypermultiplet containing
the volume is called the universal hypermultiplet, because it is
insensitive to the complex structure of $X$.

We need to give some more details.
The vector multiplet manifold is closely related to the 
K\"ahler moduli space of $X$, which is a cone, called
the K\"ahler cone. This cone is defined by the condition that the volumes of 
all curves $C \subset X$, of all holomorphic surfaces $S \subset X$, and of $X$ itself are positive when measured with the K\"ahler form $J$ 
\be
\int_{C} J > 0 \quad , \quad \int_{D} J \wedge J > 0 \quad , \quad \int_X J \wedge J \wedge J > 0 \, .
\ee
At the boundaries of the K\"ahler cone some submanifolds of $X$
contract to zero volume and $X$ becomes a singular CY space 
$\hat{X}$. This will be discussed in the next subsection. To parameterize
deformations of the K\"ahler form $J$, we introduce a basis
$C^I$, $I=0,\ldots,n_V$ for the homological two-cycles 
$H_2(X,\mathbbm{Z})$.\footnote{Recall that $n_V = h^{1,1} -1$.
The range of $I$ is chosen in order to be consistent with supergravity
conventions.} 
The dual basis $D_I$, $I=0,\ldots,n_V$ for the homological 
four-cycles $H_4(X,\mathbbm{Z})$ is defined by
\be\label{5.3}
C^I \cdot D_J = \delta^I_{~J} \,.
\ee
The K\"ahler moduli $\hat{h}^I$ are obtained by integrating the
K\"ahler form over the two-cycles $C^I$:
\be\label{5.2}
\hat{h}^I = \int_{C^I} \, J \, .
\ee
This shows that the K\"ahler moduli are in one-to-one 
correspondence with the volumes of two-cycles.\footnote{To be
precise: $\hat{h}^I$ is the minimal volume which a curve in the
homology class $C^I$ can have. This bound is saturated for
holomorphic curves.}
The volume of $X$ itself is given by
\be\label{5.6}
vol(X) = \frac{1}{3!} \int_X \, J \wedge J \wedge J = 
\ft16 C_{IJK} \hat{h}^I \hat{h}^J \hat{h}^K \,,
\ee
where we used Poincar\'e duality and the definition of 
the triple intersection numbers
\be\label{5.4}
C_{IJK} = D_I \cdot D_J \cdot D_K \;.
\ee
Since the modulus corresponding to the volume of $X$ sits
in the universal hypermultiplet, 
we need
to disentangle it from the the other K\"ahler moduli.
The rescaled fields
\be\label{5.7}
 h^I =  (6 \, vol(X))^{-1/3} \, \hat{h}^{I}\,  
\ee
are precisely the vector multiplet scalars discussed in the
last section, and the coefficients of the cubic prepotential 
${\cal V}(h)$ are precisely the triple intersection numbers.
Since the vector multiplet manifold is given by
${\cal V}(h) = 1$, we see explicitly that it is 
the projectivization of the K\"ahler cone, or, equivalently,
a hypersurface corresponding to fixed total volume.

The volume modulus $V$, which sits in the universal hypermultiplet,
is obtained by splitting the volume of $X$ into a dynamical scalar
field $V$ and  
a fixed, dimensionful `reference volume' ${\rm v}$, which 
relates the eleven-dimensional 
and the five-dimensional gravitational couplings,
\be\label{Defv}
vol (X) = {\rm v \cdot V} \;,\;\;\;\mbox{where} \;\;\;
\ft{{\rm v}}{\kappa_{(11)}^2} =  \ft{1}{\kappa_{(5)}^2} \;.
\ee
\end{subsection}

Half of the moduli in the remaining $n_H -1 = h^{2,1}$ 
hypermultiplets correspond to deformations of the complex
structure of $X$. These moduli are found by integrating
the holomorphic $(3,0)$-form of the CY manifold $X$
over the members of a basis of the homological three-cycles.
Thus, there is a one-to-one correspondence between complex
structure moduli and three-cycles. The other half of the
remaining hypermultiplet moduli correspond to deformations
of the eleven-dimensional three-form gauge field. The metric
on the hypermultiplet manifold is in general only known
at tree level. Here one can use the c-map \cite{c-map} and the whole
hypermultiplet manifold is uniquely determined by the
special K\"ahler submanifold spanned by the complex 
structure moduli of $X$. 
However, the hypermultiplet metric
is subject to perturbative and non-perturbative corrections.
Here only limited results are known, and mostly for the 
universal hypermultiplet, see \cite{AMTV,Ketov,ARV,DTV}
and references therein.

\begin{subsection}{Conifold transitions}
We now review the geometry and M-theory physics of conifold 
transitions. For a more detailed account, see
in particular \cite{GreRev,Witten1,Huebsch,Wilson,WitD=2}.

\subsubsection{Geometry: local aspects}

Conifolds are a particular type of singular CY spaces, which contain
a finite number $N$ of isolated nodes (also called double points),
which are the simplest singularities of complex threefolds. 
Let us first discuss the local geometry of a single node.
Locally a node can be described as the vertex of a cone in ${\C}^4$,
given by the zero locus of a quadratic polynomial.
One way to parameterize this cone is 
$\Psi = \sum_{a = 1}^{4} \zeta_a^2 = 0$, where 
$\zeta_a$ are complex  coordinates in ${\C}^4$. 
By intersecting the locus $\Psi = 0$ with $S^7 \subset \mathbbm{R}^8 \simeq
\mathbbm{C}^4$ one easily sees that $\Psi = 0$ is a real cone with
basis $S^2 \times S^3$. This description indicates that the singular point
can be smoothed in two different ways, by either keeping the $S^2$ or
the $S^3$ finite at the tip. The first smoothing is called the small
resolution, while the second smoothing,
where one replaces the singular point by an $S^3$, is called the
deformation.\footnote{The node can also smoothed by replacing it
by a submanifold of dimension 1,4 or 5. But this is not 
compatible  with the resulting smoothed manifold being CY.} 
The process of going from one smoothing to the other
describes the local geometry of a conifold transition:
one first tunes the K\"ahler modulus controlling a particular
two-cycle to turn it into a node. Then this modulus is frozen in,
and the node is deformed into a three-cycle, whose size is controlled
by a new complex structure modulus. Obviously, one can also  perform
this process in the opposite direction, and this is the perspective
taken in most of the literature, in particular in \cite{Str2}.
For us it will be convenient to approach the conifold singularity
`from the K\"ahler moduli  side,' because we have full control
about the vector multiplet sector, while much less is known about
the details of the hypermultiplets.

\subsubsection{Geometry: global aspects}

Next, let us consider a compact CY  space $X$ which develops
$N$ nodes when we tune the K\"ahler moduli  such that
$N$ isolated holomorphic curves ${\cal C}_i , i = 1 , \ldots , N$
shrink to points. We will denote the homology classes of these
curves by $C^{(i)}$. Naively, one might expect that
each node can be resolved or deformed independently. But this
is not true, as the condition that the resulting compact
smooth manifold is CY imposes constraints (see 
\cite{Huebsch} or \cite{Candelas2,Str2}). Generically one
can either resolve or deform all the nodes simultaneously. 
Moreover a conifold transition cannot exist if the nodes
are independent in homology. Instead, they must 
satisfies $r$ homology relations ($N > r > 0$) 
\be\label{cf_rel}
\sum_{i = 1}^N  a_{\rho i} C^{(i)} = 0 \, , \quad \rho = 1, \ldots , r .
\ee
Thus the $N$ curves which are contracted to nodes belong to 
$N-r$ different homology classes and are controlled by $N-r$ 
K\"ahler moduli. The special values of these K\"ahler moduli,
where the curves reach zero volume, belong to the boundary of
the K\"ahler cone.
The deformation of these  nodes results in a new CY 
manifold $\tilde{X}$, with Hodge numbers 
\be\label{cf.1}
h^{1,1}(\tilde{X}) = h^{1,1}(X) - (N - r) \, , \quad h^{1,2}(\tilde{X}) = h^{1,2}(X) + r \, . 
\ee
This shows that in a conifold transitions one looses $N-r$ generators
of the two-cycles and
gains $r$ generators of the three-cycles, 
or vice versa. The number of K\"ahler and
complex structure moduli changes accordingly.\footnote{While for 
the small resolution one generically has an overall
ambiguity, which leads to two different small resolutions related by
a flop, the deformation is unique:
even though the deformation contains an entire $SO(2)$-family of 
three-spheres, these are all homotopically equivalent.} 
Note that the 
Euler number changes: $\chi_{\rm E}(\tilde{X}) = \chi_{\rm E}(X) - 2 N$.

\subsubsection{Physics: microscopic aspects}

We now turn to the physical interpretation of conifold transitions
in terms of M-theory, which comprises the fundamental or
microscopic level of the description.
In the smooth CY threefold $X$ the holomorphic curves ${\cal C}_i$,
with homology classes $C^{(i)}$  can 
be expanded in the homology basis $\{ C^I \}$,
so that $C^{(i)} = q_{~I}^i C^I$, with integer coefficients $q_{~I}^i$.
Since the ${\cal C}_i$ are holomorphic, they have the minimal 
volume which is possible in their homology class,
$vol({\cal C}_i) = h^i = q_{~I}^i h^I$. 
Moreover, the $N$ cycles are subject to $r$ 
homology relations. Therefore the matrix $q_{~I}^i$ has rank 
$N-r$, and the 
$N$ volumes $h^i$ are controlled by $N-r$ of the $h^{1,1}-1$
K\"ahler moduli available at fixed total volume. 

By wrapping M2-branes around these curves, one obtains 
$N$ hypermultiplets
which carry charges $\pm (q_{~I}^i), I = 0, \ldots, n_V$, under the gauge 
fields $A^I_{\mu}$ \cite{Witten1}. 
The $i$-th
hypermultiplet couples to the particular linear combination
$A_{\mu}^{i} = q_{~I}^i A^I_{\mu}$. Since the matrix
$q_{~I}^i$ has rank $N-r$, we see that a subgroup
$U(1)^{N-r}$ of the total gauge group $U(1)^{n_V+1}$ 
is gauged.\footnote{The extra $U(1)$ is due to the graviphoton,
which belongs to the gravity multiplet.} 
Hypermultiplets are BPS multiplets, and therefore their
mass (in suitable units) is given by their central charge, which itself is
determined by the electric charges. Explicitly, the central
charge of the $i$-th hypermultiplet is $Z^{(i)} = 
q_{~I}^i h^I$. From the higher
dimensional perspective, the mass is given by the dimensional
reduction of M2-brane tension $T_{(2)}$ along the curve
${\cal C}_i$ \cite{flop}:
\be
\label{BPS_mass}
M^{(i)}_{\mscr{BPS}} 
= T_{(2)} ( 6 {\rm v} )^{1/3} \left| q_{~I}^i \, h^I \right| \;.
\ee
Here we indeed recognize the five-dimensional BPS mass
formula $M_{\mscr{BPS}}^{(i)} = \mbox{const} \cdot |Z^{(i)}|$.\footnote{
From the eleven-dimensional point of view the mass of an
M2-brane wrapped on ${\cal C}_i$ is given by $M_{(11)}^{(i)} 
= T_{(2)} vol ( {\cal C}_i)
= T_{(2)} ( 6 {\rm v V} )^{1/3} |q^i_{\;I} h^I|$. However, the 
relation between the eleven-dimensional and five-dimensional
metrics involves a conformal rescaling by the volume modulus
${\rm V}$ \cite{FS3}.} 
At the conifold point $N-r$ K\"ahler moduli are tuned 
such that $Z^{(i)} = q^i_{\;\;I}h^I = 0$ and the $N$ charged
hypermultiplets corresponding to M2-branes wrapped on the curves
${\cal C}_i$ become massless.

\subsubsection{Physics: macroscopic aspects}

Having identified the additional massless states which 
occur at the conifold point, we now address the question
how the transition can be described in terms of a five-dimensional
LEEA. Away from the conifold point, when the additional states are
heavy, we have compactifications
on smooth Calabi-Yau spaces $X$ and $\tilde{X}$. The corresponding
effective actions, $S$ and $\tilde{S}$, contain
$n_V$ and $\tilde{n}_V = n_V - (N-r)$ Abelian vector multiplets and
$n_N$ and $\tilde{n}_N = n_N - r$ neutral hypermultiplets, respectively. 
When we
approach the conifold point along the moduli space
of $X$, the $N$ charged hypermultiplets corresponding to 
M2-brane winding modes become massless. Our goal is to construct
a new, extended LEEA $\hat{S}$, which explicitly includes these
states. Thus $\hat{S}$ contains $n_V$ vector multiplets and
$n_N + N$ hypermultiplets, $N$ of which carry electric charges 
$q^i_{\;I}$, as discussed
above. By comparing to the spectrum of $\tilde{S}$, one sees
that the deformation of $\hat{X}$ into $\tilde{X}$ corresponds to a new
branch in  the moduli space along which $N-r$ vector multiplets
and $N-r$ hypermultiplets become massive. Thus there is a
Higgs effect, and the actions $S$ and $\tilde{S}$ are the effective
actions of the massless modes on the Coulomb branch and 
on the Higgs branch of $\hat{S}$,
respectively. The Higgs branch is parameterized by the scalars
of the $r$ hypermultiplets which remain massless and contain the
new additional complex structure moduli of $\tilde{X}$. 
From the analogous situation
in rigid supersymmetric theories one expects that along the
Higgs branch the $N-r$ massive vector and hypermultiplets combine
into long, non-BPS vector multiplets.\footnote{See for example
\cite{AGH}.We will see later that in our explicit gauged supergravity 
models the Higgs effect indeed works as in the rigid case.} 
The non-BPS nature of these states fits very well 
with the observation that in the microscopic picture there are
no BPS branes which could wrap three-cycles in order to give
point-like BPS states.\footnote{The exact 
microscopic origin of these degrees of freedom remains to be clarified. For the case of type IIB string theory it was argued in \cite{GreFlop} that these states correspond to massive string excitations.}

Let us next discuss how the extended LEEA $\hat{S}$ can be found
explicitly. As usual with effective actions, $S$ and $\tilde{S}$
need not be obtainable from $\hat{S}$ by simply truncating
out the additional modes. In general, integrating out states
will lead to threshold corrections, which modify some of the
terms involving the other states. This can be seen explicitly
in the vector multiplet sector, because this part of the
actions $S,\hat{S}$ and $\tilde{S}$ is determined completely
by the corresponding homogenous cubic polynomials 
${\cal V}, \hat{\cal V}$ and $\tilde{\cal V}$. For $S$ and $\tilde{S}$
these polynomials are determined by the triple intersection
numbers $C_{IJK}$ and $\tilde{C}_{\tilde{I}\tilde{J}\tilde{K}}$ 
of $X$ and $\tilde{X}$.
The polynomial $\hat{\cal V}$ can be related to ${\cal V}$
as follows: one considers the intermediate regime where the 
transition states are massive, but less heavy than all the other
massive states. Then the descriptions in terms of $\hat{S}$
and $S$ are both valid, and one can determine the relative
threshold correction between these two effective actions by
explicitly integrating out the $N$ charged hypermultiplets.
Adapting the result found in \cite{Witten1} we obtain:
\be\label{In_CIJK}
\hat{\cal V} = {\cal V} - \frac{1}{2} \, \sum_{i=1}^N \left( h^i \right)^3 \, 
= {\cal V} - \frac{1}{2} \, \sum_{i=1}^N \sum_{I=0}^{n_V} \left( 
q^i_{\;\;I} h^I \right)^3 \;.
\ee
Note that since $q^i_{\;\;I}$ has rank $N-r$, the difference
$\hat{\cal V}- {\cal V}$
only depends on the $N-r$ K\"ahler moduli which control the
$N$ curves ${\cal C}_i$.

The polynomials $\hat{\cal V}$ and $\tilde{\cal V}$ are
related by integrating out $N-r$ vector and $N-r$ hypermultiplets.
As none of these states is charged under the remaining gauge
group $U(1)^{\tilde{n}_V + 1} = U(1)^{n_V + 1 - (N-r)}$, 
there is no threshold 
correction which modifies the interactions between the remaining
vector multiplets,
and $\tilde{\cal V}$ is obtained by freezing
the $N-r$ K\"ahler moduli, which is equivalent to setting 
$h^i = 0$:\footnote{To obtain $\tilde{\cal V}$ explicitly, one
can apply a linear transformation which block-diagonalizes the charge
matrix $q^i_{\;I}$ such that the $N-r$ K\"ahler moduli controlling
the nodes correspond to, say 
$h^{I'}$ with $I'=1,\ldots, N-r$. Then one sets $h^{I'}=0$. Examples
will be given in the next two sections.}
\be
\tilde{\cal V} = \left. \hat{\cal V} \right|_{h^i=0} = 
\left. {\cal V} \right|_{h^i=0} \; .
\ee
This prescription appears to agree with the relation
between $C_{IJK}$ and $\tilde{C}_{\tilde{I} \tilde{J} \tilde{K}}$
predicted by algebraic geometry, where one drops the 
part of the intersection ring which is generated by the 
duals of the of the two-cycles $C^{(i)}$.\footnote{Examples for
the computation of triple intersection numbers and their behavior
in conifold transitions are given in Appendix C.} 
%

While the vector multiplet sector can be treated exactly, 
including threshold corrections, the situation is much less
satisfactory in the hypermultiplet sector. As explained above,
already the hypermultiplet manifolds of $S$ and $\tilde{S}$
are only known at tree level. In order to determine this manifold
for $\hat{S}$, one needs to know in addition  the couplings 
involving the groundstates of the wrapped M2-branes. 
In M-theory it is currently not known 
how to compute this.
Going to type IIA string theory by dimensional reduction does 
not help either: here one has techniques such as boundary 
conformal field theory to handle D-branes, but these can only 
be applied if the corresponding world-sheet conformal field
theory is rational, for example at Gepner points in CY
moduli spaces. However, at conifold points one does not
even have a well-defined conformal field theory, and effective
supergravity is so far the only way to see that the full, non-perturbative
string or M-theory physics is nevertheless regular. 
Because of these difficulties we will not try to determine the
exact hypermultiplet manifold from microscopic physics, but 
instead make a choice, based on simplicity, and take
the non-compact Wolf spaces $X(1+N)$ as hypermultiplet manifolds
of $\hat{S}$. As we will see this will still pose considerable
technical problems, which, however, can be solved.  Note that
we still have the essential microscopic input, namely the charge matrix
$q^i_{\;\;I}$, which determines the gauging of $\hat{S}$ 
in terms of the geometry of the transition. Since $\hat{S}$
is a gauged supergravity action, the choice of the scalar metric
and of the gauging determines the whole action, in particular
the scalar potential and therefore the mass matrix. 
Since $X(1+N)$ is a curved quaternion-K\"ahler manifold, it is 
not a priori clear whether there exists a gauging which leads
to a potential with the structure needed to describe a conifold
transition. But in the following sections we will show explicitly
that such gaugings exist and that they are uniquely determined by the
microscopic data.

\end{subsection}
\end{section}

\begin{section}{Actions for conifold transitions: a minimal model}
We now have all the ingredients to write down explicitly the LEEA for
conifold transitions, including the charged extra states which become massless
at the transition locus. To illustrate the key features of the construction,
we first consider a model with the minimal number of multiplets before turning
to the description of a general conifold transition. In both cases we proceed
by first constructing the Lagrangian and then showing that the scalar masses obey the conditions arising from the microscopic picture. 

Before embarking on the actual construction of the Lagrangian, let
us comment on the general vacuum structure expected from an effective action
describing a conifold transition. From the microscopic picture we expect
the vacuum manifold to consist of two branches, a Coulomb branch
(corresponding to the compactification on $X$) where the extra light
hypermultiplets are massive and a Higgs branch (corresponding to the
compactification on $\tilde{X}$) where some of the vector multiplets become
massive and the gauge group is spontaneously broken $U(1)^{n_V +1} \rightarrow
U(1)^{n_V +1 - (N-r)}$. 

In terms of the algebraic eqs.\ (\ref{2.24}) this
vacuum structure can be understood as follows. Suppose we gauge an isometry
where we can fix some of the hypermultiplet scalars to obtain $P^r_I(q) = 0$
while the corresponding Killing vector $K^X_I(q)$ remains non-zero. Having
$P^r_I(q) = 0$ suffices to meet the first condition in
(\ref{2.24}).\footnote{In terms of four-dimensional super-Yang-Mills theory
  the conditions $P^1_I(q) = P^2_I(q) = 0$ and $P^3_I(q) = 0$ correspond to
  $F$- and $D$-flatness, respectively \cite{AP}.} In order to satisfy the
second equation, $h^I(\phi) K^X_I(q) = 0$, we then have the choice to either
fix the remaining hypermultiplet scalars such that $K^X_I(q) = 0$ or the
vector multiplet scalars such that $h^i(\phi) =
\sum_I q^i_I h^I(\phi)= 0$.\footnote{Since the hypermultiplets are
charged with respect to $A^i_\mu = q^i_{\;\;I}A^I_\mu$, 
it is clear from (\ref{2.2}) that this is sufficient to 
have $h^I(\phi) K^X_I(q) = 0$.}
The first choice corresponds to the Coulomb branch, while the second one gives rise to the Higgs branch. The intersection locus where both $K^X_I(q) = 0$ and $h^i(\phi) = 0$ are fulfilled is the conifold locus.

\begin{subsection}{Constructing the action}
As is well known from rigid ${\cal N}=2$ supersymmetric theories, the minimal
field content for having a Coulomb and a Higgs branch is one
vector multiplet together with two charged hypermultiplets.
In the effective supergravity framework this field content can be
used to describe a conifold transition with $N=2,r=1$, if we truncate
out all the multiplets which are massless along both branches. In such
a transition the Hodge numbers are related by
$h^{1,1}(\tilde{X}) = h^{1,1}(X) -1$, 
$h^{1,2}(\tilde{X}) = h^{1,2}(X) +1$, 
while the Euler number changes according to 
$\chi_E(\tilde{X}) = \chi(\tilde{X}) - 4$. The rank of the gauge
group is reduced by one.  Besides 
the gravity multiplet, the minimal model contains only the
vector multiplet corresponding to the $U(1)$ which is Higgsed, together
with two hypermultiplets. In order to describe a conifold transition, 
we need the following behavior:
along the Coulomb branch the two hypermultiplets must be 
massive. They must become massless at the conifold 
point, where one can give a vacuum expectation value to some of 
the hypermultiplet scalars, with the result that the $U(1)$ is
Higgsed. The Higgs branch is parameterized by the scalars in one
hypermultiplet, while the vector multiplet and the other hypermultiplet
become massive.\footnote{As we will see the hypermultiplet
which parameterizes the Higgs branch is a specific linear combination of
the two hypermultiplet which we use on the Coulomb branch, 
see (\ref{11.11}), (\ref{11.11a}).}
Note that the universal hypermultiplet has been truncated out together 
with all the other neutral hypermultiplets.  
We take the vector multiplet manifold to be the
most general one, see Appendix A, while the hypermultiplet manifold
is $X(2)$.

Geometrically the conifold transition involves two 
two-cycles $C^{(i)}$, $i=1,2$, which are subject to the 
homology relation $C^{(1)} + C^{(2)} = 0$.\footnote{
We are not aware of a concrete example of a 
conifold transition with $N=2,r=1$. In Appendix
\ref{AppC} we give an example of a conifold transition
with $N=4, r=3$ which, by mirror symmetry, maps to a
conifold transition with $N=4, r=1$. The conifold transition
of the quintic has $N=16, r=1$.
The LEEA constructed in 
the next section correspond to $r=1$ with arbitrary $N$.}
The charges of the hypermultiplets 
are obtained by expanding the $C^{(i)}$ in the homology basis $C^I$, 
$C^{(1)} = - C^{(2)} = \sum_{I=0}^{n_V} q^1_{\;\;I} C^I$. The 
volume of $C^{(1)}$ is given by
$h^\star (\phi) = \sum_{I=0}^{n_V} q^1_{\;\;I} h^I(\phi)$.
In order to work with the truncated model with one vector multiplet,
we might need to perform a (not necessarily integer valued)
basis transformation of the $h^I$ such that $q^1_{\;\;I}=0$ for
$I>1$ and $h^\star (\phi) = \sum_{I=0}^1 q^1_{\;\;I} h^I(\phi)$.
The corresponding gauge field is $A^\star_\mu = \sum_{I=0}^1
q^1_{\;\;I} A^I_\mu$. The condition $C^{(1)} + C^{(2)} =0$ between
the two two-cycles translates into the requirement that the
two hypermultiplets carry opposite $U(1)$ charges.
%
%
The Killing vector encoding this charge assignment should then have the form
\be
\label{3.48b}
k_{\rm Cf}^X  =  \imag \, \left[ v^1 \, , \, - v^2 \, , \, - \vb^1 \, , \,  \vb^2 \, , \, - u_1 \, , \,  u_2 \, , \,  \ub_1 \, , \, - \ub_2 \right]^{\rm T} \, , 
\ee
with respect to the basis
\be\label{3.47a}
 \left\{ \partial_{v^1}, 
\partial_{v^2},\partial_{\vb^1},\partial_{\vb^2},\partial_{u_1},\partial_{u_2},\partial_{\ub_1},\partial_{\ub_2} \right\} \, 
.
\ee
The relative sign in the charge assignment between $v^i$ and $u_i$ stems from the fact that the $u_i$ transform in the conjugate representation of the gauge group \cite{SCgauging}.

Equating the Killing vector (\ref{3.48b}) with the expression for the most general triholomorphic Killing vector of $X(2)$, (\ref{3.42}), we find that (\ref{3.48b}) indeed is a symmetry of $X(2)$ which is generated by the Lie algebra generator 
\be\label{3.47}
t_{\rm Cf} =  {\rm diag}
\left[ 
\, 1 \, , \, -1 \, , \, 0 \, , \, 0 \, 
\right] \, . 
\ee
The $SU(2)$-triplet of moment maps associated with this isometry is then obtained by substituting this generator into the general expression for the moment maps on $X(2)$, (\ref{3.45}), and reads
\be
\label{3.49b}
\hat{\mu}^r_{\rm Cf}  =  
\left[
\begin{array}{l}
- \, \frac{\imag}{2 \phi_{+}^{1/2} \phi_{-}^{1/2}} \, \left( v^1 u_1 - v^2 u_2 - \vb^1 \ub_1 + \vb^2 \ub_2 \right) \\
- \, \frac{1}{2 \phi_{+}^{1/2} \phi_{-}^{1/2}} \, \left( v^1 u_1 - v^2 u_2 + \vb^1 \ub_1  - \vb^2 \ub_2  \right) \\
- \, \frac{1}{2 \phi_{-}} \left( u_1 \ub_1 - u_2 \ub_2 \right) + \frac{1}{2 \phi_{+}} \, \left( v^1 \vb^1 - v^2 \vb^2 \right)   
\end{array}
\right] 
\, , 
\ee
where the components of the moment map $\left\{ \hat{\mu}^1 , \hat{\mu}^2, \hat{\mu}^3 \right\}$ are adapted to the basis associated with the complex 
structures (\ref{2.7b}).

Next we need to identify the gauge connection $A^{\star}_\mu$. 
The Higgs branch condition $h^\star(\phi_c) K_{\star}^X(q_c) = 0, K_{\star}^X(q_c) \not = 0$ requires that $h^\star(\phi)$ must have a zero for some value $\phi$ inside the vector multiplet scalar manifold. The discussion in Appendix A then fixes $h^\star(\phi) = h^1(\phi)$ since $h^0(\phi)$ does not have this property. 

After determining the gauge connection $A^\star_\mu$, the Killing vector
$k_{\rm Cf}^X$,  and its associated triplet of moment maps, we now have all
the ingredients for gauging the general supergravity Lagrangian
(\ref{2.1}). We start with the kinetic terms. Since we do not gauge any
isometries of the vector multiplet scalar manifold, the corresponding gauge
covariant derivative is just the partial derivative,
\be\label{4.5}
\cD_{\mu} \phi^x = \partial_\mu \phi^x \quad \Leftrightarrow \quad K^x_{I}(\phi) \quad \mbox{\rm 
trivial} \, .
\ee
We now turn to the gauging of the hypermultiplet sector. Since the
$U(1)$  gauge connection of the isometry (\ref{3.48b}) is $A^\star_\mu =
A^1_\mu$ we must set
\be\label{4.6}
 K^X_0(q) = 0  \, , \qquad K^X_1(q) = k_{\rm Cf}^X(q) \,  .  
\ee
The covariant derivative for the hypermultiplet scalars then becomes
\be\label{4.7}
\cD_\mu \, q^X = \partial_\mu \, q^X + {\rm g} \, A^1_\mu \, k_{\rm Cf}^X(q) \,
. 
\ee
This  expression explicitly shows that the two hypermultiplets $v^1, u_1$ and $v^2, u_2$ carry opposite $U(1)$ charge with respect to the gauge connection $A^1_\mu$. Furthermore, 
we have included an explicit gauge coupling constant ${\rm g}$.
At the level of five-dimensional 
gauged supergravity this is a free parameter, but 
we will see later that ${\rm g}$ is fixed by the microscopic 
data of the full M-theory.

Next we turn to the scalar potential (\ref{2.20}). Taking into account the relations (\ref{3.51}) we set
\be
P^r_0(q) = 0 \, , \qquad P^r_1(q) = \frac{1}{2} \, \hat{\mu}^r_{\rm Cf}(q) \; .
\ee
Correspondingly, $P^r$ is obtained as
\be
P^r(\phi, q) = \frac{1}{2} \, h^1(\phi) \, \hat{\mu}^r_{\rm Cf}(q) \, .
\ee
In order to construct the scalar potential, we work out the superpotential (\ref{2.21}). For the $P^r$ above this is given by
\be\label{11.10}
\begin{split}
\cW = & \frac{ h^1 }{\sqrt{6}} \, \bigg\{ \left( \frac{1}{2 \phi_{+}} \left( v^1 \vb^1 - v^2 \vb^2 \right) - \frac{1}{2 \phi_{-}} \left( u_1 \ub_1 - u_2 \ub_2 \right) \right)^2 \\
& \qquad \qquad + \frac{1}{\phi_{+} \phi_{-}} \left( v^1 u_1 - v^2 u_2 \right) \left( \vb^1 \ub_1  - \vb^2 \ub_2 \right) \bigg\}^{1/2} \, .
\end{split}
\ee
From this expression it is then straightforward to check that $Q^r$ defined in
 eq.\ (\ref{2.22}) is independent of the vector multiplet scalar. This implies
 that the scalar potential $\pV(\phi, q)$ is determined by 
the `superpotential' ${\cal W}$, because it can be brought to  
the stability form (\ref{2.23}). 
\end{subsection}
\begin{subsection}{Vacua and mass matrix}
The supersymmetric vacua of this potential are determined by the conditions (\ref{2.24}). Looking at the moment map (\ref{3.49b}) reveals that $\hat{\mu}^r_{\rm Cf}$ vanishes if and only if $v^1 = v^2 , u_1 = u_2$. Restricting the Killing vector (\ref{3.48b}) to this subset we find that the resulting expression is non-vanishing. The second condition in (\ref{2.24}) can then be met by either setting the hypermultiplet scalars to zero, which corresponds to the Coulomb branch
\be\label{11.11}
\cM_{\rm Coul} = \left\{
v^1 = u_1 = v^2 = u_2 = 0 \, , \, \phi \; \mbox{undetermined}  
 \right\} \, ,
\ee
or fixing $\phi = 0$ which yields the Higgs branch
\be\label{11.11a}
 \cM_{\rm Higgs} = \left\{
v^1 = v^2 =: v \, , \, u_1 = u_2 =: u \, , \, \phi = 0 \, \right\} \, .
\ee 
On the latter, the non-trivial vacuum expectation value of the hypermultiplet scalars spontaneously breaks the $U(1)$ gauge group. Restricting the superpotential (\ref{11.10}) to these vacua, we find that the resulting expression vanishes. This implies that these vacua are Minkowski, establishing that our model has the correct vacuum structure to describe a conifold transition. 

To complete our discussion, we also calculate the masses of the fields on the two vacuum branches. For the scalars the masses are given by the eigenvalues of the mass matrix
\be\label{4.19}
\cM^\Lambda_{~~\Sigma} = {\rm g}^2 \, \left.  g^{\Lambda \Xi} \, \frac{\partial}{\partial \phi^\Xi} \, \frac{\partial}{\partial 
\phi^\Sigma} \, \pV(\phi, q) \right|_{\cM_{\rm Vac} } \, ,
\ee
where $g^{\Lambda \Sigma}$ has been defined in eq.\ (\ref{4.14}). Arranging the coordinates on $\cM_{\rm HM} \times \cM_{\rm VM}$ as
\be\label{11.12a}
\phi^{\Lambda} = \left\{ \,  v^1 \, , \, v^2 \, , \, \vb^1 \, , \, \vb^2 \, , \, u_1 \, , \, u_2 \,  , \, \ub_1 \, , \,  \ub_2 \, , \, \phi \, \right\} \, ,
\ee
the evaluation of the mass matrix on the Coulomb branch  yields
\be\label{11.12}
\left. \cM^{\Lambda}_{~\Sigma} \right|_{\rm Coul} = (m_{\rm Coul})^2 \, {\rm diag}\left[ \, 1 \, , \, 1 \, , \, 1 \, , \, 1 \, , \, 1 \, , \, 1 \, , \, 1 \, , \, 1 \, , 0 \, \right] \, ,
\ee
with
\be\label{11.13}
\left( m_{\rm Coul} \right)^2 = \frac{3}{2} \,  {\rm g}^2 \, \left( h^1 \right)^2 \, . 
\ee
In terms of the microscopic picture $|h^1|$ corresponds to the volume of the
shrinking cycle. This implies that (\ref{11.13}) has precisely the structure 
expected from the eleven-dimensional point of view. By comparing
with (\ref{BPS_mass}) and using (\ref{Defv}) together with
the value $T_{(2)} = ( \ft{8 \pi}{\kappa_{(11)}^2})^{1/3}$ 
of the M2-brane tension \cite{flop}, we find that ${\rm g}$ is
fixed by M-theory,
\be
{\rm g} = \sqrt{\ft23} (48 \, \pi)^{1/3} \;.
\label{4.23}
\ee
Thus the extended LEEA is completely fixed once we
choose the hypermultiplet manifold to be $X(2)$.

Let us now turn to the Higgs branch. Here we have to evaluate the mass matrix
for the scalar fields (\ref{4.19}) and in addition obtain a mass term for the
vector field $A^1_\mu$. In the latter case we note that since $a_{IJ}|_{\phi =
  0} = \unit_{2 \times 2}$ (cf. Appendix A), the kinetic 
term for the vector fields has its standard form, so that the mass of $A^1_\mu$ can be read off directly from the term proportional to $\left( A^1_\mu \right)^2$ arising in the hypermultiplet kinetic term. Defining
\be\label{11.14}
\left( m_{\rm Higgs} \right)^2 = 2 \, {\rm g}^2 \, \left( \frac{1}{\phi_{+}} \, v \vb  + \frac{1}{\phi_{-}} u \ub \right) \, , 
\ee
where $\phi_\pm$ is understood to be restricted to $\cM_{\rm Higgs}$, we find
that the vector field $A^1_\mu$, the vector multiplet scalar $\phi$ and three
hypermultiplet degrees of freedom acquire the mass $m_{\rm Higgs}$ while the
remaining fields stay massless. In terms of
supersymmetry algebra representations (see for example \cite{AGH}) these modes
organize into one massless hypermultiplet and one long vector multiplet of
mass $m_{\rm Higgs}$. The latter contains 
all the massive degrees of freedom and one massless hypermultiplet mode 
which is the would-be Goldstone boson that 
provides the longitudinal mode of the vector field. 
This shows explicitly that the vector multiplet `eats up' one of the 
hypermultiplets and becomes a non-BPS multiplet. 
\end{subsection}
\end{section}
\begin{section}{Actions for conifold transitions: general case}
In the last section we illustrated the key features of an extended LEEA 
describing a conifold
transition by means of a minimal model.
Now we generalize this setup to the case where $N$ two-cycles satisfying $r=1$ homology relations contract to nodes.  When including the neutral universal hypermultiplet, the hypermultiplet sector of the corresponding Lagrangian contains $N+1$ hypermultiplets and thus will be modeled by $X(N+1)$. In order to calculate the masses of the fields on the Higgs branch  we  need to specify an explicit vector multiplet sector. Here we will use the family of very special real manifolds discussed in Appendix \ref{appF} including an arbitrary number $n_V \ge N-1$ of vector multiplets. Even though these manifolds are not directly related to an exact 
vector multiplet sector arising from a CY compactification, this ansatz is general enough that it can readily be adjusted to any particular conifold transition by substituting the corresponding vector multiplet sector obtained from eq. (\ref{In_CIJK}).
\begin{subsection}{Constructing the action} 
The construction of the Lagrangian describing a general conifold transition proceeds exactly as in the case of the minimal model. Our first task is to generalize the single Killing vector (\ref{3.48b}) to a set of Killing vectors $k^X_\alpha(q)$ which encode the charges of the extra hypermultiplets with respect to the gauge fields $A^\alpha_\mu$ associated with the $h^\alpha(\phi)$ parameterizing the volume of the collapsing cycles ${C}^{(\alpha)}$. The microscopic analysis from Subsection 2.3 indicates that the charges are encoded in a set of Killing vectors $ \alpha = 1, \ldots , N-1$ of the form
\be\label{11.1}
\begin{split}
\hat{k}^v_\alpha  & = \imag \left[ 0 \, , \, v^2 \, , \,  0 , \ldots , \, - v^{\alpha +2} \,  , 0 , \ldots, 0 \right]^{\rm T} \, , \\
\hat{k}^u_\alpha  & = \imag \left[ 0 \, , \, - u^2 \, , \,  0 , \ldots , \, u^{\alpha +2} \, , 0 , \ldots, 0 \right]^{\rm T} \, , 
\end{split}
\ee 
where the entries `$-v^{\alpha +2} $' and `$u^{\alpha +2} $' occur at the
$\alpha +2$ position. The components of $\hat{k}^X_\alpha$ with respect to
$\partial_{\vb^i}$ and $\partial_{\ub_i}$ are obtained from
$\hat{k}^{v}_\alpha$ and $\hat{k}^u_\alpha$ by complex conjugation. Here we
have chosen our hypermultiplets such that $v^1, u_1$ parameterize the universal
hypermultiplet, $v^2, u_2$ arise from the M2-brane wrapping the cycle
${C}^{(1)} = - \sum_{\alpha =1}^{N-1} {C}^{(\alpha+1)} $, and the 
remaining hypermultiplets $v^{\alpha +2}, u_{\alpha +2}$ correspond to the M2-branes wrapping the cycles ${C}^{(\alpha+1)} $, $\alpha  = 1, \ldots, N-1$, respectively. The set of Killing vectors (\ref{11.1}) mutually commutes, $\left[\hat{k}_{\alpha} , \hat{k}_{\beta} \right] = 0$, so that the hypermultiplets are charged with respect to a subgroup $U(1)^{N-1} \subset U(1)^{n_V+1}$. Note that the minimal model of the last section is obtained from (\ref{11.1}) by truncating the universal hypermultiplet (erasing the first slot in the Killing vectors) and restricting to the case $N = 2$.

Our next task is to determine the generators $t_\alpha \in su(N+1,2)$ which give rise to the isometries (\ref{11.1}). For this purpose we first restrict the expression for a general Killing vector on $X(N+1)$, (\ref{3.42}), to the subset of Killing vectors which are linear in the hypermultiplet scalar fields
\be\label{6.2}
\begin{split}
\hat{k}^{v^i} & = \imag \left( t^i_{~j} v^j - v^i \, t^{n+1}_{~~~n+1} \right) \, ,  \\
\hat{k}^{u_i} & = \imag \left( t^{n+1}_{~~~n+1} u_i - t^j_{~i} u_j - u_i \left( t^{n+1}_{~~~n+1} - t^{n}_{~~n} \right) 
\right) \, .
\end{split}
\ee
Equating this expression with the Killing vectors (\ref{11.1}) and imposing the condition that $t_\alpha$ is traceless leads to an overdetermined system of equations from which the generators $t_\alpha$ can be determined. In the case of the Killing vectors (\ref{11.1}) we find the solution 
\be\label{11.21}
t_\alpha = {\rm diag} \, \left[ \, 0 \, , \, 1 \, , \, 0 \, , \, \ldots \, , \, -1 \, , \, 0 \, , \, \ldots \, , \, 0 \, \right] \, ,
\ee
where the entry `$-1$' sits at the $\alpha+2$ position. Hence the vectors (\ref{11.1}) are indeed tri-holomorphic isometries of $X(N+1)$.

In the next step we calculate the moment maps $\hat{\mu}^r_\alpha$ of these isometries. Substituting the generators $t_\alpha$ into eq. (\ref{3.45}) and taking the linear combinations $\hat{\mu}^1 = \hat{\mu}^+ + \hat{\mu}^-$, $\hat{\mu}^2 = - \imag \left( \hat{\mu}^+ - \hat{\mu}^- \right)$ we obtain 
\be\label{11.22}
\hat{\mu}^r_\alpha  =  
\left[
\begin{array}{l}
- \, \frac{\imag}{2 \phi_{+}^{1/2} \phi_{-}^{1/2}} \, \left( v^2 u_2 - v^{\alpha + 2} u_{\alpha + 2} - \vb^2 \ub_2 + \vb^{\alpha + 2} \ub_{\alpha + 2} \right) \\
- \, \frac{1}{2 \phi_{+}^{1/2} \phi_{-}^{1/2}} \, \left( v^2 u_2 - v^{\alpha+2} u_{\alpha + 2} + \vb^2 \ub_2  - \vb^{\alpha+2} \ub_{\alpha+2}  \right) \\
- \, \frac{1}{2 \phi_{-}} \left( u_2 \ub_2 - u_{\alpha + 2} \ub_{\alpha + 2} \right) + \frac{1}{2 \phi_{+}} \, \left( v^2 \vb^2 - v^{\alpha + 2} \vb^{\alpha + 2} \right)   
\end{array}
\right] 
\, . 
\ee

In order to complete the construction of the Lagrangian, we then have to gauge these isometries. Following our previous discussion we need $N-1$ gauge fields $A^\alpha_\mu$ whose corresponding scalar fields $h^\alpha(\phi)$ vanish at some subset of the vector multiplet scalar manifold. By virtue of eq. (\ref{E.5}) this condition is satisfied for all $h^I(\phi)$, $I \geq 1$, 
but not for $h^0(\phi)$. Without loss of generality we then choose the gauge fields $A^\alpha_\mu$, $\alpha = 1, \ldots , N-1$, to serve as the gauge connections for the Killing vectors.

The gauging of the Lagrangian proceeds analogous to the gauging in the minimal 
model. Again we start by considering the kinetic terms. As in the previous case, we do not gauge any isometries in the vector multiplet sector, so that the corresponding covariant derivative is again given by eq.\ (\ref{4.5}). In the hypermultiplet sector we gauge the subgroup $U(1)^{N-1} \subset U(1)^{n_V}$. With the choice of gauge connections made above, the covariant derivative in the hypermultiplet kinetic term becomes
\be\label{4.7a}
\cD_\mu \, q^X = \partial_\mu \, q^X + {\rm g} \, A^\alpha_\mu \, k_\alpha^X(q) \,
. 
\ee
This expression indicates that the universal hypermultiplet $v^1, u_1$ is neutral, while the hypermultiplet scalar field $v^2$ carries charge $q = +1$ with respect to all gauge fields $A^\alpha_\mu, \alpha = 1, \ldots, N-1$ and the fields $v^{\alpha+2}$ have charge $q = -1$ with respect to $A^\alpha_\mu$ only. This charge assignment then is in complete agreement with the microscopic description of the transition.

Finally we construct the scalar potential (\ref{2.20}) of our Lagrangian. Evaluating the expressions (\ref{2.19}) for the particular gauging at hand we obtain
\be\label{11.23}
P^r = \frac{1}{2} \, h^\alpha(\phi) \, \hat{\mu}_{\alpha}^r(q) \, , \quad P^r_x = \frac{1}{2} \, h^\alpha_x(\phi) \, \hat{\mu}^r_\alpha(q) \, , \quad K^X(\phi, q) = h^{\alpha}(\phi) \, k^X_{\alpha}(q) \, ,
\ee
with the sets $k^X_\alpha(q)$ and $\hat{\mu}_\alpha^r(q)$ given in (\ref{11.1}) and (\ref{11.22}), respectively.
The scalar potential of this Lagrangian is determined by substituting these relations, the hypermultiplet metric (\ref{3.31}) and the inverse vector multiplet scalar metric (\ref{E.7}) into
\be\label{11.24}
\pV(\phi, q) = -4 P^r P^r + 2 g^{xy} P^r_x P^r_y  + \frac{3}{4} g_{XY} 
K^X K^Y \;.
\ee
This completes the construction of the Lagrangian. Let us remark that rewriting $\pV(\phi, q)$ in terms of the superpotential (\ref{2.21}) is not helpful in this case as it is simpler to determine the vacuum structure and mass matrices of the Lagrangian in terms of the moment maps and the potential (\ref{11.24}).
\end{subsection}
\begin{subsection}{Vacua and mass matrices}
We now check whether the supersymmetric vacuum structure and the mass matrices of the Lagrangians constructed above agree with the specifications coming from the microscopic theory. Determining the vacuum structure again utilizes the algebraic eqs. (\ref{2.24}) and proceeds completely analogous to the minimal model. Looking at the set of moment maps (\ref{11.22}) reveals that they vanish if and only if all charged hypermultiplets have the same value, i.e., if $v^2 = \ldots = v^{N+1} = v$ and $u_2 = \ldots = u_{N+1} = u$. To solve the second condition in (\ref{2.24}) we then have the choice to either fix $v=u=0$, leading to the Coulomb branch
\be\label{11.25}
\cM_{\rm Coul} = \left\{
v^2 = \ldots = v^{N+1} = 0 \, , u_2 = \ldots = u_{N+1} = 0 \, 
 \right\} \, ,
\ee
or to set $\phi^\alpha = 0 \Leftrightarrow h^\alpha(\phi) = 0$ which corresponds to the Higgs branch
\be\label{11.26}
 \cM_{\rm Higgs} = \left\{
v = v^2 = \ldots = v^{N+1} \, , \, u = u_2 = \ldots = u_{N+1} \, , \, \phi^\alpha = 0 \, \right\} \, .
\ee 
The scalars not present in these equations are not fixed by the vacuum conditions and correspond to flat directions. We further note that the non-trivial value of the hypermultiplet scalars on the Higgs branch breaks the $U(1)^{N-1} \subset U(1)^{n_V + 1}$ spontaneously.
Substituting the conditions of vanishing $\hat{\mu}^r_ \alpha$ and $K^X$ into the potential (\ref{11.24}) shows that these vacua are Minkowski so that we obtain the correct vacuum structure expected for a conifold transition.

Our next task is to calculate the mass matrices for these vacuum branches.  We first analyze the Coulomb branch before turning to the Higgs branch.

In the first step of determining the mass matrix (\ref{4.19}) we check which terms of the potential (\ref{11.24}) give rise to non-zero contribution. Here the first observation is that for the $P^r$, (\ref{11.23}), the terms $P^r P^r$ and $g^{xy} P^r_x P^r_y $ are of 
fourth order in the extra hypermultiplets. This implies that these terms do not contribute to the mass matrix on the Coulomb branch since 
they vanish identically when taking two derivatives with respect to any scalar field and restricting to $\cM_{\rm Coul}$ afterwards. 
Hence  the masses of our fields are solely generated by the last term in eq. (\ref{11.24}). 

In the next step we show that the vector multiplet scalar fields $\phi^x$ are massless. The matrix
\be\label{6.14}
\cM_{\Lambda \Sigma} := \left. \partial_\Lambda \partial_\Sigma \left( 
\frac{3}{4} \, {\rm g}^2 \, g_{XY} \,  K^X \, K^Y 
\right) \right|_{\cM_{\rm Coul}}
\ee
has non-trivial entries if and only if both $\Lambda$ and $\Sigma$ take values in the hypermultiplet sector. To see this, we 
expand $K^X = h^{\alpha}(\phi) k^X_{\alpha}(q)$ and note that $k^X_{\alpha}(q)$ vanishes when restricted to $\cM_{\rm Coul} $. 
This implies $\cM_{\Lambda \Sigma}$ is only non-trivial if there is one derivative acting on each of the Killing vectors $k^X_{\rm \alpha}(q)$.
Since $g^{\Xi \Lambda} =  g^{XY} \oplus g^{xy}$ is the 
direct sum of the hypermultiplet and vector multiplet inverse metrics, we find that  non-trivial entries of the mass matrix 
(\ref{4.19}) may occur in the hypermultiplet sector only. This establishes that the vector multiplet scalars $\phi^x$ are 
massless. In order to determine the masses of the hypermultiplet scalar fields we then restrict ourselves to evaluating the matrix 
\be\label{11.27}
\cM_{XY} = \left. \frac{3}{2} \, {\rm g}^2 \, h^\alpha \, h^\beta \,  g_{WZ} \, \left( \partial_X \, k^W_{\alpha} \right) \, \left( \partial_Y \, k^Z_{\beta} \right) \, \right|_{\cM_{\rm Coul}} \, .
\ee
For this purpose we first compute $K^Y_{~X} := h^\alpha \partial_X k^Y_\alpha(q)$. Using the basis 
\be\label{6.18}
q^X = \left\{ \, v^1 \,, \ldots  , v^{N+1} \, , \, \vb^{1} \, , \ldots  , \, \vb^{N+1} \, , \, u_1 \, , \ldots  , \, u_{N+1} 
\, , \, \ub_1 \, , \ldots  , \, \ub_{N+1} \,  \right\} 
\ee
the resulting matrix is diagonal,
\be\label{11.28}
K^Y_{~X} = \imag \, {\rm diag} \, \left[ \, H \, , \, -H \, , \, -H \, , \, H \,  \right] \, ,
\ee
with 
\be\label{11.29}
H = {\rm diag} \left[ \, 0 \, , \, - \sum h^\alpha \, , \, h^1 \, , \, \ldots \, , \, h^{N-1} \, \right] \, .
\ee

Next we restrict the hypermultiplet scalar metric $g_{XY}(q)$ to the Coulomb branch. We find that all blocks appearing in $g_{XY}(q) |_{\cM_{\rm Coul} }$ are diagonal:
\be\label{6.20}
\begin{array}{ll}
g_{v_1 \vb_1}  =  \frac{1}{2 \phi_+^2 \phi_-^2} \left( 1 - \ub_1 u_1  \left( 1 - \vb^1 v^1 \right)^2 \right)  \, , \quad &
 g_{v^{\alpha+1} \vb^{\beta+1}}   =   \frac{1}{2 \phi_+} \, \delta_{\alpha+1,  
\beta +1 } \, , \\
g_{u_1 \ub_1}   =  \frac{1}{2 \phi_-^2} \left( 1 - \vb^1 v^1 \right)  \,  , &
g_{u_{\alpha +1}  \ub_{\beta+1}}  =   \frac{1}{2 \phi_-} \, 
\delta^{\alpha +1 , \beta +1}  \, ,
\\
g_{\vb^1 u_1}  =   - \,  \frac{1}{2 \phi_-^2} \left( \ub_1 v^1 \right) \, . & 
\end{array}
\ee
Here and in the following $\phi_+$ and $\phi_-$ are understood to be restricted to $\cM_{\rm Coul} $. The matrix $\cM_{XY}$ can now be 
computed from 
\be\label{6.21}
\cM_{XY} = \, \frac{3}{2} {\rm g}^2 \, \left[ K \, g \, K^{\rm T} \right] \, .
\ee
Explicitly, we find
\be\label{6.22}
\cM_{XY} =  \, 
\left[ 
\begin{array}{cccc}
 0 & A & 0 & 0 \\
 A & 0 & 0 & 0 \\
 0 & 0 & 0 & B \\
 0 & 0 & B & 0 \\
\end{array}
\right] \, ,
\ee
with $A$ and $B$ being the following $(N+1) \times (N+1)$-dimensional block matrices:
\be\label{11.30}
\begin{split}
A = & \, \frac{3}{4 \phi_+} \, {\rm g}^2 \, \left[ \, 0 \, , \, \left( \sum h^\alpha \right)^2 \, , \, \left( h^1 \right)^2 \, , \, \ldots \, , \, \left( h^{N-1} \right)^2  \right] \, , \\
B = & \, \frac{3}{4 \phi_-} \, {\rm g}^2 \, \left[ \, 0 \, , \, \left( \sum h^\alpha \right)^2 \, , \, \left( h^1 \right)^2 \, , \, \ldots \, , \, \left( h^{N-1} \right)^2  \right] \, . 
\end{split}
\ee

Finally we need to calculate the inverse metric $g^{XY}$, restricted to $\cM_{\rm Coul} $, by inverting $g_{XY} |_{\cM_{\rm Coul} }$ given in 
(\ref{6.20}).  The only non-zero components of $g^{XY}|_{\cM_{\rm Coul} }$ are given by
\be\label{6.24}
\begin{array}{ll} 
  g^{v^1 \vb^1}    =   2 \, \phi_+^2 \, \phi_-  \, , \quad &
 g^{v^{\alpha +1} \vb^{\beta + 1}}   =   2 \phi_+ \delta^{\alpha + 1,  
\beta + 1} \, , \\
 g^{u_1 \ub_1 }  =  2 \, \frac{\phi_-}{\phi_+} \left( 1 - \ub_1 u_1 \left( 1 - \vb^1 v^1 \right)^2 \right) \, , \quad & 
g^{u_{\alpha +1} \ub_{\beta + 1}}   =  2 \, \phi_- \, 
\delta^{\alpha + 1 ,  \beta + 1} \, , \\
g^{v^1 \ub_1}   =  2 \phi_+ \, \phi_- \left( \ub_1 v^1 \right) \, .  &
\end{array}
\ee

The hypermultiplet masses are given by the eigenvalues of the mass matrix
\be\label{6.25}
\cM^X_{~~Y} = \left.g^{XZ} \, \cM_{ZY} \right|_{\cM_{\rm Coul} } \, .
\ee
 Explicitly we find the masses for the universal hypermultiplet $v^1,u_1$, the
 hypermultiplet $v^2,u_2$ arising from the wrapped cycle 
${C}^{(1)} = - \sum_{\alpha=1}^{N-1} {C}^{(\alpha+1)}$ and the other
 hypermultiplets $v^{\alpha+2}, u_{\alpha+2}$ associated with the cycles 
${\cal C}^{(\alpha+1)}$, $\alpha = 1,\ldots,N-1$, of volume $h^{\alpha}$ to be
\be\label{11.31}
m_1 = 0 \, , \quad m_2 = \sqrt{\frac{3}{2}} \, {\rm g} \, \sum_{\alpha =
  1}^{N-1} h^\alpha \, , \quad \mbox{and} \quad 
m_{\alpha+2} = \sqrt{\frac{3}{2}} \, {\rm g} \, h^{\alpha} \, ,  
\ee
respectively. These masses are in complete agreement with the microscopic theory, which demands that the mass of the hypermultiplets is proportional to the volume of the cycle wrapped by the corresponding M2-brane. Comparing these masses to the BPS mass formula (\ref{BPS_mass}) fixes ${\rm g}$ to the value (\ref{4.23}). This input pins down the remaining freedom of the extended LEEA in terms of the microscopic data.

Let us now turn to the Higgs branch. Here we do not have a prediction for the masses of the massive states in terms of the microscopic description. But since the data on the Coulomb branch together with the charge assignment (\ref{11.1}) completely determines the Lagrangian, we can use our model to predict the masses on this branch.

We again begin our calculation by investigating which terms of the potential (\ref{11.24}) contribute to the mass matrix on the Higgs branch. The first term proportional to $P^r P^r$ is of second order in both $\phi^\alpha$ and $\hat{\mu}_\alpha^r$ which both vanish when restricted to the Higgs branch. Hence all its second derivatives vanish when restricted to $\cM_{\rm Higgs}$ and there is no contribution to the mass matrix. The second term, $2 g^{xy} P^r_x P^r_y$, has orders zero and two in $\phi^\alpha$ and $\hat{\mu}^r_\alpha$, respectively. This indicates that this term is of relevance for the hypermultiplet mass matrix, as the only non-vanishing second derivative terms (restricted to $\cM_{\rm Higgs}$) are obtained when one derivative with respect to a hypermultiplet scalar acts on each of the moment maps. Finally the third term is of second order in $\phi^\alpha$ while the Killing vectors $k^X_\alpha$ do not vanish on $\cM_{\rm Higgs}$. Therefore this term gives rise to the vector multiplet scalar masses, as we have to take two derivatives with respect to the vector multiplet scalars to have a non-vanishing expression when restricting to $\cM_{\rm Higgs}$.

The analysis of the supersymmetry representations \cite{AGH} and the explicit example of our minimal Lagrangian show that the massive modes on this branch combine into massive long vector multiplets composed of one vector and one hypermultiplet of the same mass.
%
%
In order to determine this mass it suffices to compute the vector multiplet scalar masses arising from the third term in the potential.

We start by evaluating
%
\be\label{11.32}
\cM_{xy} = \left. \frac{3}{2} \, {\rm g}^2 \, g_{XY} \, \Big( \partial_x \, h^\alpha(\phi) \, k^X_{\alpha} \Big) \, \Big( \partial_y \, h^\beta(\phi) \, k^Y_{\beta} \Big) \, \right|_{\cM_{\rm Higgs}} \, .
\ee
To this end we first calculate the $(n_V) \times 4 ( N+1 )$ dimensional matrix $K_x^{~X} := \partial_x \, h^\alpha(\phi) \, k^X_{\alpha}$. Taking the basis (\ref{6.18}) for the hypermultiplets and using the definition (\ref{E.5}) for $h^I(\phi)$ we find
\be\label{11.33}
K_x^{~X} = h^0 \left[ 
\begin{array}{cccc}
- \imag \, v^2 \, \cQ &  \imag \, \vb^2 \, \cQ &  \imag \, u_2 \, \cQ & - \imag \, \ub_2 \, \cQ \\
          0       &            0          &            0        &            0
\end{array}
\right] \, , 
\ee
where $\cQ$ denotes the $(N-1) \times (N+1)$ matrix\footnote{This matrix is the charge matrix of the $N+1$ hypermultiplets with respect to the $U(1)^{N-1}$ gauge connections $A^\alpha_\mu$.}
\be\label{11.34}
\cQ = \left[ 
\begin{array}{cc|c}
0 & -1 & \\
\vdots & \vdots & \quad \unit_{(N-1) \times (N-1)} \\
0 & -1 & 
\end{array}
\right] \, ,
\ee
and `$0$' is short for the $(n_V - N + 1) \times (N+1)$ dimensional zero matrix.

It is then straightforward to calculate 
$\cM_{xy} =  \frac{3}{2} \, {\rm g}^2 \, \left[ K \, g \, K^{\rm T}\right]_{xy} \, \big|_{\cM_{\rm Higgs}}$ 
to be
\be\label{11.35}
\cM_{xy} = 3 \, {\rm g}^2 \, (h^0)^2 \, {\rm diag} \left[ \, \Pi \, , \, 0 \, \right] \, .  
\ee
Here $\Pi$ denotes the $(N-1) \times (N-1)$ matrix
\be\label{11.36}
\Pi = |v|^2 \, \cQ g_{v \vb} \cQ^{\rm T} + |u|^2 \, \cQ g_{u \ub} \cQ^{\rm T} - v \ub \, \cQ g_{\ub v} \cQ^{\rm T} - u \vb \, \cQ g_{u \vb} \cQ^{\rm T} \, ,   
\ee
where $v,u$ are the values of the hypermultiplet scalars on the Higgs branch and $g_{\vb v}$, $g_{\ub u}$, $g_{\ub v}$ and $g_{u \vb}$ denote the block matrices appearing in the hypermultiplet scalar metric $g_{XY}(q)$. Introducing $M := \cQ g \cQ^{\rm T}$ with $g$ representing the blocks of $g_{XY}(q)$ we find that there is a simple relation between the components of $M$ and $g$
\be\label{11.37}
m_{\alpha \beta} = g_{22} - g_{(\alpha+2) 2} - g_{2 (\beta+2)} + g_{(\alpha+2) (\beta+2)} \, . 
\ee
Evaluating this expression on $\cM_{\rm Higgs}$, we find $M=0$ if $g = g_{u \vb}$ or $g = g_{\ub v}$ while for $g=g_{v \vb}$ and $g = g_{u \ub}$ we obtain 
\be\label{11.38}
m_{\alpha \beta} = \frac{1}{2 \phi_+} \, \left( 1 + \delta_{\alpha \beta} \right) \quad \mbox{and} \quad 
m_{\alpha \beta} = \frac{1}{2 \phi_-} \, \left( 1 + \delta_{\alpha \beta} \right) \, , 
\ee
respectively. This allows us to write down $\cM_{xy}$ explicitly. Its non-zero components read
\be\label{11.38a}
\cM_{\alpha \beta} = \frac{3}{2} \, {\rm g}^2 \, (h^0)^2 \, \left( \frac{1}{\phi_+} \, |v|^2 + \frac{1}{\phi_-} \, |u|^2 \right) \, \big( 1 + \delta_{\alpha \beta} \big) \, . 
\ee

The mass matrix $\cM^x_{~y}$ for the vector multiplet scalars is obtained by raising one index of $\cM_{xy}$ using the inverse vector multiplet scalar metric restricted to the Higgs branch (\ref{E.8}). Defining
\be\label{11.39}
\left( m_{\rm Higgs} \right)^2 := {\rm g}^2 \, \bigg( (h^0)^2 - \frac{3}{2} \, \sum_{\sigma=N}^{n_V} (h^{\sigma})^2 \bigg) \, \bigg( \frac{1}{\phi_+} |v|^2 + \frac{1}{\phi_-} \, |u|^2 \bigg) \, , 
\ee
where $\phi_+$ and $\phi_-$ are understood to be restricted to $\cM_{\rm Higgs}$, we find
\be\label{11.42}
\cM^x_{~y} = \left( m_{\rm Higgs} \right)^2 \, {\rm diag} \left[ \cM^\alpha_{~ \beta} \, , \, 0 \, \right] \, , 
\ee
with $\cM^\alpha_{~ \beta} = 1 + \delta^{\alpha}_{~ \beta}$.

This result shows that all `spectator vector multiplets' $h^\sigma, \sigma = N, \ldots , n_V$, which do not serve as a gauge connection for the isometries remain massless. Determining the eigenvalues of $\cM^\alpha_{~ \beta}$ shows that the vector multiplet scalars $\phi^\alpha$ acquire two different masses. One scalar has  mass $(N)^{1/2} \, m_{\rm Higgs}$ while the remaining $N-2$ scalars have degenerate mass $m_{\rm Higgs}$. Here it is interesting to observe that, in the presence of spectator vector multiplets, the mass of these states depends on both the vector and hypermultiplet degrees of freedom. This 
points at the non-BPS nature of the corresponding long vector multiplets. We also observe that in the absence of spectator vector multiplets and for $N=2$ the general formula (\ref{11.42}) agrees with the results obtained for the minimal Lagrangian of the previous subsection.

In terms of CY compactifications, these general Lagrangians have the following interpretation. On the Coulomb branch, associated with the CY compactification on $X$, we have $n_V$ vector and one hypermultiplet which are generically massless. At the conifold locus one obtains $N$ additional massless hypermultiplets. When going to the Higgs branch arising from the CY compactification on $\tilde{X}$, $N-1$ charged hypermultiplets are `eaten up' by the massive vector multiplets to form long vector multiplets. We are then left with $n_V - (N - 1)$ vector and two hypermultiplets which are generically massless. This establishes that our 
extended LEEA indeed provides a continuous description of a conifold
transition where the Hodge numbers change by $h^{1,1}(\tilde{X}) =
h^{1,1}(X)-(N-1)$ and $h^{1,2}(\tilde{X}) = h^{1,2}(X)+1$, while
the Euler number changes by $\chi_E(\tilde{X}) = \chi_E(X)-2N$.
In particular, this class includes the conifold transition of the quintic studied in \cite{Str2,Pol2} where the Hodge numbers of $X$ and $\tilde{X}$ are given by $h^{1,1}(X) = 102$, $h^{1,2}(X) = 0$ and $h^{1,1}(\tilde{X}) = 87$, $h^{1,2}(\tilde{X}) = 1$. 
\end{subsection}  
\end{section}

\begin{section}{Conclusions}

In this paper we have constructed gauged supergravity actions which
model the LEEA of M-theory compactified on a CY threefold
undergoing a conifold transition. These actions explicitly include 
the extra states descending from wrapped
M2-branes and have a non-trivial scalar potential with
a Coulomb and a Higgs branch. While their vector multiplet sector was determined exactly,
we made a particular choice for the hypermultiplet scalar metrics in order to 
obtain explicit actions.
In a companion paper \cite{CC} we 
use these actions to investigate the M-theory physics
of conifold transitions, in particular the
dynamical realization of conifold transitions and 
the moduli  dynamics in cosmological solutions.

Since the explicit actions constructed in this paper
only cover the case where an arbitrary number of two-cycles satisfying $r=1$ homology relations are contracted, more work has to be done to extend our construction to the case of general $r$. Ultimately one would of
course like to derive the metrics on the scalar 
manifolds from M-theory.  Presumably, the exact effective
action $\hat{S}$ is determined by combining geometrical
data of the singular space $\hat{X}$ with data 
involving the M2-branes wrapping the vanishing cycles.
The extended LEEA found in this paper suggest that
this combined set of data is meaningful 
and in some sense smooth. 

To explain what we mean let us elaborate
on the discussion of the relation between threshold corrections
and geometry in \cite{MohZag}.
Recall that the
vector multiplet sector can be treated exactly\footnote{
In Section 4 we restricted the vector multiplet manifold to 
be of a special type, but only in order to be able to compute
the mass matrix explicitly.} and assigns coefficients
$C_{IJK}$, $\hat{C}_{IJK}$, $\tilde{C}_{\tilde{I} \tilde{J} \tilde{K}}$
to the spaces $X,\hat{X},\tilde{X}$. Since $C_{IJK}, 
\tilde{C}_{\tilde{I} \tilde{J} \tilde{K}}$ are the triple
intersection numbers of $X$ and $\tilde{X}$, we expect that
$\hat{C}_{IJK}$ can be interpreted as triple intersection numbers
of the singular space $\hat{X}$. Also note that the combined
web of moduli spaces of CY manifolds $X,\tilde{X}$ related
by conifold transitions is not a manifold, but only a 
stratification, because its dimension jumps at the transition 
point. 
We can, however, construct the extended LEEA $\hat{X}$, associated with the geometrical quantities of the singular space $\hat{X}$ to obtain a smooth description of the physics at the transition locus where the actions $S$ and $\tilde{S}$ are discontinuous.
%
Thus it seems that M-theory implicitly gives us a prescription
of how to resolve the intersection points of the web of
moduli spaces, and how to assign regular data to the corresponding
singular spaces. This observation is not restricted to conifold
transitions, but also applies to flop transitions as well as 
extremal transition \cite{MohZag} which involve non-Abelian gauge symmetry
enhancement. It would be very interesting to make
these ideas mathematically precise.

Another possible extension of our work is to consider
the four-dimensional LEEA of type II string theories
on CY threefolds. While essentially nothing changes in 
the hypermultiplet sector, the vector multiplet sector
gets more complicated, because it is now governed
by a holomorphic prepotential, instead of a real cubic
one. Moreover, the threshold corrections 
of the vector multiplet couplings resulting from integrating
out charged multiplets show a different behavior in five
and in four dimensions. As we have seen in Section 2,
the five-dimensional threshold correction only induce discontinuities
in the third derivatives of the prepotential. In four dimensions,
however, second derivatives of the prepotential (gauge couplings)
become logarithmically divergent as a function of the modulus
controlling the distance to the conifold point \cite{Str1}, while
its third derivatives (the `Yukawa couplings') have a first order
pole. However, we expect that it is possible to handle
these threshold corrections along the lines of \cite{LouMohZag}, where
it was shown that one can `integrate in' four-dimensional
vector multiplets. It would be interesting to 
extend this work to the case where one also integrates in 
hypermultiplets. 
Furthermore, conifold transitions are special  cases of 
extremal transitions \cite{Extremal}. In the full class one also
finds transitions where one does not have isolated
singular points, but curves of singularities. 
This leads to extra massless charged vector multiplets, and,
hence, to the un-Higgsing of non-Abelian gauge groups, which could also be phenomenologically interesting.

Another line of research extending the present setup would be 
the inclusion of background fluxes.
Fluxes could in particular 
lift the still existing flat directions and break the 
still existing supersymmetry spontaneously. A particular 
interesting idea is that the interplay between flux and
vanishing cycles can generate a hierarchically small 
scale, and thus might help to solve the gauge hierarchy 
and the cosmological constant problem \cite{Hierarchy}. It is also
known that when fluxes a turned on, the remaining vacua 
prefer to sit at special points in moduli space, such as
conifold points \cite{Str1,FluxSpecial}. Such points could be investigated by 
including fluxes into our framework. 

Also note that the 
deformed and the resolved conifold geometry can be obtained,
at least locally, as solutions of gauged supergravity 
theories \cite{GaugedConifold}, 
where the gauging results from wrapping branes on cycles or
switching on background fluxes. These gaugings are different
from those considered in this paper: they do not correspond to 
integrating in charged multiplets, they are non-compact 
rather than compact, they lift directions in scalar field 
space, which  in our setup are necessarily flat, and they can
break supersymmetry spontaneously. This reflects the different
roles played by branes: while in \cite{GaugedConifold} branes (or the flux 
they resolve into) are part of the background, we have included
certain light brane degrees of freedom as dynamical degrees
of freedom into the Lagrangian. It would be interesting to 
better understand the relation between these complementary
setups.

Finally, we would like to step outside the framework
of type II CY compactifications (and their deformations by fluxes)
and investigate topological phase transitions in 
genuine $\cN=1$ compactifications such as CY compactifications
of type I and heterotic strings, heterotic M-theory,
type II CY orientifolds
and M-theory compactifications on sevenfolds with $G_2$ 
structure.\footnote{See for example \cite{AchGuk} for a recent
review of the role of singularities and topology change
in $G_2$ compactifications 
of M-theory.} Here topological phase transitions, for
example conifold transitions and small instanton transitions,
can change the chiral matter content \cite{ChiralTransitions}. 
Part of the 
LEEA technology applies to these cases as well.
In particular the computation of threshold corrections 
resulting from  integrating out states is straightforward,
at least at the one loop level, see for example \cite{CLM}.
Of course, when considering LEEA with $\cN=1$ supersymmetry
one seems to loose the powerful tool of special
geometry. However, there are indications that in
string compactifications some features of special
geometry survive, so that at least holomorphic quantities
are accessible \cite{SpecN=1}.

\end{section}

\begin{section}*{Acknowledgments}
This work is supported by the DFG within the `Schwerpunktprogramm 
String\-theorie'. F.S.\ acknowledges a scholarship from the `Studienstiftung
des deu\-tschen Volkes' and FOM. We thank Sakura Schafer-Nameki for useful 
discussions about the CFT-related aspects of type II conifold transitions. 
\end{section}

\begin{appendix}
\begin{section}{Vector multiplet sector I: one multiplet}
\label{appA}
The simplest extended LEEA describing a conifold transition contains two charged hypermultiplets and was constructed in Section 4. The vector multiplet sector of this action consists of a single vector multiplet. For convenience we collect  all the relevant properties of this sector in this appendix.

Following \cite{EGZ}, the prepotential (\ref{2.4}) of the most general one vector multiplet scalar manifold can be brought to the standard form
\be\label{B.1}
{\cal V} = (h^0)^3 - \frac{3}{2} \, h^0 \, (h^1)^2 + c \, (h^1)^3 = 1 \, ,
\ee
where $c$ denotes an arbitrary real constant. The value $c = - \frac{1}{\sqrt{2}}$ corresponds to the manifold $O(1,1)$ considered in \cite{UHM} while the prepotential
\be\label{11.1a}
{\cal V} = \frac{3}{8} U^3 + \frac{1}{2} U \, T^2 = 1 \, , 
\ee
which arises from Calabi-Yau compactifications on $T^2 \times K3$ \cite{Jan,phasetransitions} can be mapped to the case $|c| > \frac{1}{\sqrt{2}}$.  

The metric coupling the vector field kinetic terms is obtained from eq.\ (\ref{2.5}) and reads
\be\label{B.3} 
a_{IJ} = \left[ 
\begin{array}{cc}
a_{00} & a_{01} \\ 
a_{10} & a_{11} 
\end{array}
 \right] \, ,
\ee
with entries
\bea\label{B.3a}
\nonumber
a_{00} & = &  -2 h^0 + 3 \, \left( (h^0)^2 - \frac{1}{2} \, (h^1)^2 \right)^2 \, , \\ 
a_{01} & = & h^1 - \frac{3 \, h^1}{2} \, \left(2 \, (h^0)^2 - (h^1)^2 \right) \, \left(h^0 - c \, h^1 \right) \, ,  \\ \nonumber
a_{11} & = & h^0 - 2\, c \, h^1 + 3 \, \left( h^0 \, h^1 - c \, (h^1)^2 \right)^2 \, .
\eea
Observe that at $h^{0} = 1$, $h^1 = 0$, which we interprete as 
a conifold point in Section 3, 
this becomes the two-by-two unit matrix. Hence the two vector field kinetic terms decouple and have standard normalization at this point.

To fully specify the action, we also need a parameterization of the vector multiplet scalar manifold. Here it is convenient to introduce the vector multiplet scalar field $\phi$ as 
coordinate
\be
\phi = \frac{h^1}{h^0} \;. 
\ee
Solving this relation for $h^1$ and substituting the resulting expression into the prepotential (\ref{B.1}), it is straightforward to calculate $h^I(\phi)$ introduced in eq.\ (\ref{2.4}):
\be\label{B.2}
h^0(\phi) = \left( 1 - \frac{3}{2} \, \phi^2 + c \, \phi^3  \right)^{-1/3} \, , \quad h^1(\phi) = \phi \left( 1 - \frac{3}{2} \, \phi^2 + c \, \phi^3  \right)^{-1/3}.
\ee 
The vector multiplet scalar manifold (\ref{2.7}) is then given by 
\be\label{B.4}
g_{xy} = \frac{3 \left( 2 - 4 \, c \, \phi + \phi^2 \right)}{\left( 2 - 3 \, \phi^2 + 2 \, c \, \phi^3 \right)^2} \, .
\ee 

At the conifold point $h^1(\phi) = 0 \leftrightarrow \phi = 0$ this metric is regular $g_{xy}|_{\phi = 0} = \frac{3}{2}$. The coordinate range of $\phi$ is bounded by the values $\phi_{-}^{\rm crit}$ and $\phi_{+}^{\rm crit}$, at which the metric becomes zero or infinite. For $|c| < \frac{1}{\sqrt{2}}$ the metric diverges at both $\phi_{-}^{\rm crit}$ and $\phi_{+}^{\rm crit}$ while for $|c| \ge \frac{1}{\sqrt{2}}$ we encounter one zero and one infinity bounding the coordinate range of $\phi$. For the special case $c = 0$ considered in 
\cite{CC} we have $\phi^{\rm crit}_{\pm} = \pm \sqrt{\frac{2}{3}}$ implying the corresponding vector multiplet scalar manifold is given by the interval $\left] - \sqrt{\frac{2}{3}}  , \sqrt{\frac{2}{3}}  \right[$. Note that even though the coordinate length of this interval is finite, the boundaries $\phi^{\rm crit}_{\pm}$ are at infinite geodesic distance from the origin. 
\end{section}
\begin{section}{Vector multiplet sector II: $n_V$ multiplets}
\label{appF}
The construction of the extended LEEA describing  a general conifold transition in Section 4 required specifying an explicit vector multiplet sector containing $n_V$ vector multiplets. In this appendix we derive the properties of the vector fields and vector multiplet scalar metrics needed in this construction. The calculation is analogous to the one vector multiplet case considered in Appendix \ref{appA}, but due to the increased number of vector multiplets somewhat more involved.

We start by specifying the prepotential (\ref{2.4}) to be
\be\label{E.1}
{\cal V} = (h^0)^3 - \frac{3}{2} \, h^0 \, \sum_{x=1}^{n_{V}} (h^x)^2 = 1 \, ,
\ee
where $x,y,z = 1, \ldots , n_V$ counts the vector multiplets. The prepotential is again given in the standard form \cite{EGZ}, but is not the most general one, as we have chosen $C_{xyz} = 0$ in order to keep the calculation manageable.

The metric coupling the kinetic terms of the vector fields is obtained from eq.\ (\ref{2.5})
\be\label{E.2} 
a_{IJ} = \left[ 
\begin{array}{cc}
a_{00} & a_{0x} \\ 
a_{x0} & a_{xy} 
\end{array}
 \right] \, ,
\ee
with the entries being
\bea\label{E.3}
\nonumber
a_{00} & = &  -2 h^0 + 3 \, \left( (h^0)^2 - \frac{1}{2} \, \sum_{z=1}^{n_{V}} \, (h^z)^2 \right)^2 \, , \\ 
a_{0x} & = & h^x - 3 \, h^0 \, h^x \, \left( (h^0)^2 - \frac{1}{2} \, \sum_{z=1}^{n_V} (h^z)^2 \right) \, ,  \\ \nonumber
a_{xy} & = & h^0 \, \delta_{xy} + 3 \, (h^0)^2 \, h^x \, h^y \, .
\eea
In order to have an explicit parameterization of the vector multiplet scalar manifold we introduce homogeneous coordinates
\be\label{E.4}
\phi^x = \frac{h^x}{h^0} \; , \quad x = 1, \ldots , n_{V} \, . 
\ee
Solving these relations for $h^x$ and substituting the resulting expressions into the prepotential (\ref{E.1}) the $h^I$ become functions of the vector multiplet scalars $\phi^x$: 
\be\label{E.5}
h^0(\phi) = \bigg( 1 - \frac{3}{2} \, \sum_{z=1}^{n_{V}} ( \phi^z )^2  \bigg)^{-1/3} ,  \; \;  h^x(\phi) = \phi^x \, \bigg( 1 - \frac{3}{2} \, \sum_{z=1}^{n_{V}}(\phi^z)^2 \bigg)^{-1/3} .
\ee 
These relations are used to calculate the components of the vector multiplet scalar metric (\ref{2.7})
\be\label{E.6}
g_{xy} = \left\{ 
\begin{array}{ll}
\frac{ 3 \, \left( 1 + \frac{1}{2} (\phi^x)^2 - \frac{3}{2} \sum^{n_{V}}_{z \not=x} (\phi^z)^2 \right)}{2 \, \left( 1 - \frac{3}{2} \sum_{z=1}^{n_{V}} (\phi^z)^2 \right)^2} \, , 
& \mbox{for} \; x = y \, , \\[1.5ex]
\frac{6 \, \phi^x \, \phi^y}{\left( 1 - \frac{3}{2} \sum_{z=1}^{n_{V}} (\phi^z)^2 \right)^2} \, , 
& \mbox{for} \; x \not= y \, . 
\end{array}
\right. 
\ee
In order to find the mass matrix on the Higgs branch we further need the inverse vector multiplet scalar metric $g^{xy}$ whose components are given by
\be\label{E.7}
g^{xy} = \left\{ 
\begin{array}{ll}
\frac{2 \, \left( 1 - \frac{3}{2} (\phi^x)^2 + \frac{1}{2} \sum_{z \not=x}^{n_{V}} (\phi^z)^2  \right)  \, \left( 1 - \frac{3}{2} \sum^{n_{V}}_{z = 1} (\phi^z)^2 \right)}{3 \, \left( 1 + \frac{1}{2} \sum_{z=1}^{n_{V}} (\phi^z)^2 \right)} \, ,
& \mbox{for} \; x = y \, , \\[1.5ex]
- \, \frac{ 4\, \phi^x \, \phi^y \, \left( 1 - \frac{3}{2} \sum_{z=1}^{n_{V}} (\phi^z)^2 \right) }{3 \left( 1 + \frac{1}{2} \sum_{z=1}^{n_{V}} (\phi^z)^2 \right) } \, , & \mbox{for} \; x \not= y \, . 
\end{array}
\right. 
\ee

The vacuum conditions on the Higgs branch generically fix a subset $\phi^\alpha, \alpha = 1 , \ldots , N-r \le n_V$, to be zero (cf.\ eq.\ (\ref{11.26}) for the case $r=1$). By virtue of the relation (\ref{E.5}) this corresponds to the $h^\alpha$ parameterizing the volume of the contracting cycles being zero. Restricting $g^{xy}$ given above to $\phi^\alpha = 0$, the resulting expression becomes block diagonal 
\be\label{E.8}
g^{xy} \big|_{\phi^\alpha = 0} = \left[ 
\begin{array}{cc}
a \, \unit_{(N-r) \times (N-r)} & 0 \\
0 & g \\
\end{array}
\right] \, , 
\ee
where
\be\label{E.9}
a := \frac{2}{3} \, \left( 1 - \frac{3}{2} \, \sum_{\sigma = m + 1}^{n_{V}} (\phi^{\sigma})^2 \right) \, ,
\ee
and $g$ is to be understood as $g^{\sigma_1 \sigma_2} |_{\phi^\alpha = 0}$ 
with $\sigma_1,\sigma_2 = N-r+1, \ldots , n_{V}$. This data completely determines the vector multiplet sector of the conifold model discussed in Section 4 and enables us to explicitly calculate the vector multiplet scalar masses on the Higgs branch.
\end{section}
\begin{section}{An explicit conifold transition}
\label{AppC}
In this appendix, we give the complete geometrical data for a particular pair of CY manifolds $X$ and $\tilde{X}$ which are related by a conifold transition.  The discussion follows \cite{Huebsch} where further details can be found.

Our sample CY manifolds $X,\tilde{X}$ are so-called complete intersecting Calabi-Yau manifolds (CICY) \cite{CICY,Huebsch}. These are special in the sense that they can be obtained as complex submanifolds of an embedding space ${\mathbf \chi}$ which is given as the product of $m$ complex projective spaces ${\mathbbm P}^{n_r}_r$, $r = 1, \ldots, m$,
\be\label{C.1}
{\mathbf \chi} = {\mathbbm P}^{n_1}_1 \times \ldots \times {\mathbbm P}^{n_m}_m \, . 
\ee
The submanifold is defined by the intersection of the zero loci of $K$ polynomials, which are homogeneous with respect to the coordinates of the different ${\mathbbm P}^{n_m}_m$-factors. Generally CICY are characterized by a configuration matrix
\be\label{C.2}
{\cM} \in 
\left[ 
\begin{array}{c||c}
\bf{n} & \bf{q}
\end{array}
\right] 
= 
\left[ 
\begin{array}{c||ccc}
n_1 & q^1_1 & \ldots & q^1_K \\
\vdots & \vdots &  & \vdots \\
n_m & q^m_1 & \ldots & q^m_K  
\end{array}
\right]
\ee
with the $n_r$ being the dimension of ${\mathbbm P}^{n_r}_r$ in (\ref{C.1}) and the integers $q_a^r$, $a = 1, \ldots, K$ denoting the degree of the $K$-th constraint polynomial with respect to the homogeneous coordinates of 
${\mathbbm P}^{n_r}_r$.

Introducing $J_r$ as the K\"ahler form of ${\mathbbm P}^{n_r}_r$ the Euler characteristic of the manifold $\cM$ can be computed by evaluating the relation \cite{Huebsch}
\be\label{C.3}
\left[ \frac{\prod^m_{r=1} (1 + J_r)^{n_r+1}}{\prod^K_{a = 1} \left( 1 + \sum_{r=1}^m q^r_a \, J_r \right)} \right] \cdot \prod_{a = 1}^{K} \left( \sum_{r = 1}^m q^r_a \, J_r \right) = \ldots + \chi_{\rm E}(\cM) \cdot \prod^m_{r = 1} \left( J_r \right)^{n_r} 
\ee
by expanding the LHS in $J_r$ and evaluating the coefficient of the term $\prod^m_{r = 1} \left( J_r \right)^{n_r}$. 

As a concrete example we consider the families of CY $X$ and $\tilde{X}$ given by the configurations 
\be\label{C.4}
X \in \left[ 
\begin{array}{c||ccc}
5 & 4 & 1 & 1 \\
1 & 0 & 1 & 1  
\end{array}
\right]_{-168}^{2}
\, , \quad
\tilde{X} \in \left[ 
\begin{array}{c||cc}
5 & 4 & 2\\
\end{array}
\right]_{-176}^{1} \, . 
\ee
Here the subscript denotes the Euler number $\chi_E$ of the configuration computed from eq.\ (\ref{C.3}) while the superscript gives its Hodge number $h^{1,1}$ which is taken from \cite{Huebsch}.

For the configuration $X$ the embedding space is ${\mathbf \chi} = {\mathbbm P}^{5}_1 \times {\mathbbm P}^{1}_2$. Taking the homogeneous coordinates on ${\mathbf \chi}$ and ${\mathbbm P}^{5}_1 \times {\mathbbm P}^{1}_2$ to be $x^\alpha$, $\alpha = 1, \ldots ,6$ and $y^\beta$, $\beta = 1,2$ the three constraint polynomials can be written as 

\be\label{C.5}
Q(x)  =  0 \; , \qquad
y^1 \, x^1 + y^2 \, x^2  =  0  \; , \qquad
y^1 \, x^3 + y^2 \, x^4  =  0 \, , 
\ee
where $Q(x)$ denotes some quadric polynomial in the coordinates $x^\alpha$. Explicitly, we will use $Q(x) = \sum_{\alpha = 1}^6 (x^\alpha)^4$ for which (\ref{C.5}) becomes a set of transverse polynomials. Let us now consider the last two constraint equations in (\ref{C.5}). As $y^1$ and $y^2$ are homogeneous coordinates in ${\mathbbm P}^{1}_2$, they cannot vanish simultaneously. Satisfying these equations then requires that the determinant
\be\label{C.6}
C(x) = x^1 \, x^4 - x^2 \, x^3 = 0 
\ee
vanishes. This constraint has a double zero at the points $x^1 = x^2 = x^3 = x^4 = 0$ indicating that the gradient 
vanishes at these points, ${\rmd C}(x) = 0$. This shows that if we shrink 
the radius of the $P^1_2$ in 
${\mathbf \chi} = {\mathbbm P}^{5}_1 \times {\mathbbm P}^{1}_2$ in (\ref{C.5}) to zero we obtain a singular CY space $\hat{X}$. Substituting $x^1 = x^2 = x^3 = x^4 = 0$ into the first constraint $Q(x) = 0$ 
\be\label{C.7}
Q(x) = \sum_{\alpha=1}^6 (x^\alpha)^4 \, , \quad \left( 0 , 0, 0, 0, \omega, 1 \right)
\ee
we find that $\hat{X}$ develops four isolated nodes corresponding to the four solutions of $\omega^4 = -1$. Hence, by shrinking the ${\mathbbm P}^{1}_2$ we obtain a singular CY $\hat{X}$ where $N = 4$ holomorphic curves have collapsed to points. 

These nodes can be smoothed by adding a quadratic polynomial $t \cdot r(x)$ to (\ref{C.6}) which does not vanish at the nodes of $\tilde{X}$. The corresponding desingularized CY manifold is  given by the zero locus of the quadric and quadratic polynomials 
\be\label{C.8}
Q(x) = 0 \, , \quad C(x) + t \, r(x) = 0 \, ,
\ee
in the embedding space ${\mathbf \chi} = {\mathbbm P}^{5}_1$, corresponding to the configuration $\tilde{X}$ in (\ref{C.4}).\footnote{In the mathematical language $X$ and $\tilde{X}$ are said to be related by a operation called determinant splitting \cite{CICY,connected2} which connects all CICY \cite{connected,connected2}.} 

From the discussion above, we see that the moduli spaces of $X$ and
$\tilde{X}$ share some common points where $X$ and $\tilde{X}$ develop
nodes. The difference in the Euler characteristics (\ref{C.4}) indicates that
the topological transition connecting $X$ and $\tilde{X}$ has to be a conifold
transition. 
Substituting the Hodge numbers $h^{1,1}(X) = 2$ and $h^{1,1}(\tilde{X}) = 1$
into the relation (\ref{cf.1}) we obtain, that the $N = 4$ contracting curves
${\cal C}_i$ (which are in the same homology class) satisfy $r  = 3$ homology
relations. Hence the configurations (\ref{C.4}) are related by a conifold
transition with $N = 4$, $r = 3$. Since mirror symmetry exchanges the roles of 
even-dimensional and odd-dimensional cycles, and of 
K\"ahler and complex structure moduli, the corresponding
mirror manifolds are related by a conifold transition with 
$N=4$, $r=1$.  

Our particular example further has the advantage that it is fairly simple to calculate the triple intersection numbers $C_{IJK}$ of the configurations (\ref{C.4}). In this context we note that the Lefschetz hyperplane theorem (see \cite{Huebsch} for details) applies to $X$, $\tilde{X}$. This guarantees that the $H^{1,1}$-cohomology is generated by the pullback of the K\"ahler forms $J_r$ of the ${\mathbbm P}^{n_r}_r$. The K\"ahler form 
$J_r \in H^{1,1}({\mathbbm P}^{n_r}_r)$ may then be represented by adding a linear constraint on ${\mathbbm P}^{n_r}_r$ to the configuration which is independent of the other coordinates in the other factors in ${\mathbf \chi}$. 
In fact, it can be seen from eq.\ (\ref{C.3}) that adding a linear constraint to a configuration (\ref{C.2}) results in an additional factor $J_r$ to the LHS product in (\ref{C.3}), so that the triple intersection numbers\footnote{Note that the factor $\prod_{a = 1}^{K} \left( \sum_{r = 1}^m q^r_a \, J_r \right)$ works like a delta function, restricting the integration on the embedding space ${\mathbf \chi}$ to the submanifold $\cM$.} 
\be\label{C.9}
C_{rst} = \int_{\cM} J_r \, J_s \, J_t = \int_{\mathbf \chi} J_r \, J_s \, J_t \wedge \prod_{a = 1}^{K} \left( \sum_{r = 1}^m q^r_a \, J_r \right) 
\ee
can be obtained by adding three linear constraints to the configuration matrix (\ref{C.2}) and then evaluating (\ref{C.3}) for this extended configuration. For the particular configuration $X$ we represent the constraints corresponding to $J_1$ and $J_2$ by
\be\label{C.10}
X_0 := \left[ \begin{array}{c} 1 \\ 0 \end{array} \right] \; , \qquad X_1 := \left[ \begin{array}{c} 0 \\ 1 \end{array} \right] \, ,  
\ee
respectively. For $\tilde{X}$ the constraint $X_1$ drops out as this constraint is associated with the factor ${\mathbbm P}^{1}_2$ which vanishes on $\tilde{X}$.

Thus the $C_{IJK}$ of the configurations $X$ and $\tilde{X}$ are then be computed as
\be\label{C.11}
C_{IJK} = \chi_{E} \left[ \begin{array}{c||cccc} {\mathbf \chi} & {\bf q} & X_I & X_J & X_K \end{array} \right] \, .
\ee
Explicitly, we find
\be\label{C.12}
\begin{split}
C_{000} & =  8 \, , \quad C_{001} = 4 \, , \quad C_{011} = 0 \, , \quad C_{111} = 0 \, , \\
\tilde{C}_{000} & =  8 \, .
\end{split}
\ee
This implies that the prepotentials of $X$ and $\tilde{X}$ are given by
\bea\label{C.13}
\cV & = & 8 \, (h^0)^3 + 12 \, (h^0)^2 \, (h^1) = 1 \, , \, \mbox{and} \\ \label{C.14}
\tilde{\cV} & = & 8 \, (h^0)^3 = 1 \, ,  
\eea
respectively. Here we see that $\cV$ and $\tilde{\cV}$ are indeed related by the truncation rule given in Subsection 2.3 which, in the case at hand corresponds to setting the volume of the shrinking cycle, $h^1$, to zero. 
%
%
\end{section}
\end{appendix}

\end{document}